%% file: main.tex
\documentclass{sig-alternate-10pt}
\usepackage[margin=1in]{geometry}

\usepackage{graphicx}

\usepackage{xspace}
\usepackage{url}
\usepackage{paralist}
\usepackage{multirow}
\usepackage{array}
\usepackage[center]{subfigure}

\begin{document}



\hyphenation{proj-ects}

\title{From BGP to RTT and Beyond:\\Matching BGP Routing Changes and
Network Delay Variations with an Eye on Traceroute Paths\thanks{Supported by EU FP7
Project ``Leone: From Global Measurements to Local Management'', no.\ 317647. 
}
}

\author{Massimo Rimondini\\\affaddr{Roma Tre University}\\\email{rimondin@dia.uniroma3.it}\and
        Claudio Squarcella\\\affaddr{Roma Tre University}\\\email{squarcel@dia.uniroma3.it}\and
        Giuseppe {Di~Battista}\\\affaddr{Roma Tre University}\\\email{gdb@dia.uniroma3.it}
}

\maketitle

\begin{abstract}
Many organizations have the mission of assessing the quality
of broadband access services offered by Internet Service Providers (ISPs). They
deploy network probes that periodically perform network measures towards selected Internet services.
By analyzing the data collected by the probes it is often possible to gain a reasonable
estimate of the bandwidth made available by the ISP. However, it is much more difficult to
use such data to explain who is responsible of the fluctuations of other network
qualities. This is especially true for latency, that is fundamental for several nowadays
network services. On the other hand, there are many publicly accessible BGP routers that
collect the history of routing changes and that are good candidates to be used
for understanding if latency fluctuations depend on interdomain routing.

In this paper we provide a methodology that, given a probe that is located inside the
network of an ISP and that executes latency measures and given a set of publicly
accessible BGP routers located inside the same ISP, decides which routers are best
candidates (if any) for studying the relationship between variations of network
performance recorded by the probe and interdomain routing changes.
We validate the methodology with experimental studies based on data gathered by the RIPE
NCC, an organization that is well-known to be independent and that publishes both BGP
data within the Routing Information Service (RIS) and probe measurement data within
the Atlas project.
\end{abstract}

\section{Introduction}

The perceived quality of Internet services strongly depends on the performance of the
network portion that customer traffic has to traverse in order to reach them. With the
goal of improving this quality, in the last few years many organizations have started to
massively assess the performance of the network in reaching specific targets. For example,
the RIPE Atlas probes~\cite{atlas} continuously perform Pings, Traceroutes, and HTTP
requests towards selected destinations like root name servers and suitably located Web
sites. Although these organizations gather huge amounts of performance information,
analyzing the collected data is a complex task, which is often accomplished with limited
accuracy or completely left out. This is especially true when the target service and the
probes are located in distinct Internet Service Providers (ISPs). 

Measured performance are subject to fluctuations that depend on many networking factors,
including: bandwidth, network congestion, and routing changes. 
In this paper we focus on analyzing the relationship between variations of network
performance and interdomain routing changes. More precisely, we consider the following
quite common scenario, depicted in Fig.~\ref{fig:reference-scenario}. A set $P$ of probes
is deployed within the network of a certain ISP. These probes perform periodical RTT and
router-level path measurements (pings, traceroutes) towards a set $T$ of targets that are
spread anywhere in the Internet, corresponding to relevant services. The results of these
measurements are stored for later use. A set $C$ of BGP routers, called Collector Peers
(CPs), is also deployed within the same ISP. These routers have peerings with BGP routers
of neighboring ISPs and collect and store BGP updates received through these peerings. We
highlight that in this scenario different CPs may collect at the same time different
information on the routing towards the same target. Similarly, different probes may record
different performance in reaching the same target.
Such a scenario is nowadays quite common: projects like~\cite{samknows,atlas,bismark} have
a growing base of deployed probes, while other projects like~\cite{ris,route} provide
historical BGP routing data collected from hundreds of CPs.

The infrastructure described above is a valuable source of information for the kind of
analysis we are interested in. In particular, we consider the following specific problem:
given a probe in $P$ and a target in $T$, which CPs in $C$ are best candidates for
studying the relationship between variations of network performance recorded by the probe
and interdomain routing changes?
Besides its methodological importance, this investigation is relevant from an applicative
point of view: in fact, organizations that assess the quality of consumer broadband
services can gain a much deeper insight in the quality of the access offered by an ISP,
rather than relying only on an estimate of the available bandwidth.
In this investigation we concentrate our attention on RTT measurements, because this is
the most commonly available type of measurement, ICMP echo request packets towards
Internet services are unlikely to be filtered out by routers, and the importance of
latency in nowadays services is increasing, as pointed out by several authors (see,
e.g.,~\cite{Lee:2012:MSA:2398776.2398788}).

The problem we tackle is as difficult as ``finding a needle in a haystack'' for several
reasons: we do not assume any knowledge on the network topology; CPs may be deployed at
arbitrary locations, and may not coincide with egress points for traffic exiting the ISP;
the BGP updates collected by CPs only determine the forward path to the target, while
information on the reverse path is not available; last, routing does not consist of the
sole interdomain part, and CPs do not have any information about intradomain routing
changes occurring in the traversed Autonomous Systems (ASes).

Despite these difficulties, we introduce an effective methodology for finding the
``needle,'' namely for addressing the above stated problem. The methodology takes as input
a sequence of RTT measurements collected by a probe and a sequence of BGP updates
collected by a CP, and produces as output a value in $[0,1]$ that determines the amount of
correlation between the two sequences. The methodology is based on the following main
steps:
\begin{inparaenum}[(a)]
   \item detection of significant value changes in the sequence of RTT measurements, based on a statistical technique called PELT~\cite{kfe-odcwlcc-12};
   \item compensation of possible misalignments between the time instants recorded for RTT measurements and those recorded for BGP updates; and
   \item matching of RTT variations and BGP updates, to determine the amount of correlation.
\end{inparaenum}
We apply our methodology extensively to publicly available data sets from the RIPE Routing
Information Service~\cite{ris} and RIPE Atlas~\cite{atlas}, and validate its results using
analytical methods and by comparison with a ground truth derived from traceroute data
collected by the RIPE Atlas probes.

The rest of the paper is organized as follows. In Section~\ref{sec:relatedwork} we review
the state of the art on projects devoted to assessing the quality of service offered by
ISPs and on methods to investigate the relationships between control plane information and
measurement information. Section~\ref{sec:reference-scenario} is devoted to describe our
reference scenario. In Section~\ref{sec:methodology} we illustrate our
correlation methodology. In Section~\ref{sec:experiments} we validate the results
obtained by applying our methodology to publicly available data. Conclusions and future work are presented in Section~\ref{sec:conclusions}.

\input{020-relatedwork}

\input{025-scenario}

\input{030-methodology}

\input{050-experiments}

\input{055-validation}

\section{Conclusions and Future Work}\label{sec:conclusions}
In this paper we describe a methodology for analyzing the relationship between variations of network performance measured by probes and interdomain routing changes recorded by BGP route collectors. We show several examples of correlation that can be discovered between the two kinds of data by using our methodology. Moreover, we discuss how to tune the input parameters of our methodology for best results and validate it using traceroute information collected by the probes as a ground truth.

There are lots of facets of our methodology that deserve further investigation.
First of all, at present we consider actual RTT values only to compute the changepoint analysis, and actual BGP paths only during the validation. Actual traceroute paths are not fully considered in the validation phase either. We believe the quality of the correlation could be further improved by taking these into account also at other steps of the methodology.
Moreover, at present we discard spurious RTT measurements: instead, they could be possible hints of the presence of a routing problem or change.
A solid criterion to automate the decision whether RTT data for a \textsf{Target} and BGP data for a \textsf{Prefix} have a good correlation is still to be determined.
We could also further investigate the impact of methodology parameters on the produced results, for example by considering different functions for determining the penalty for the \textsc{Changepoint detection} or different width and centering for the \textsf{Tolerance window}.
In addition, we should take into account possible changes in the location of the probes over time.
The validation process could finally be enhanced by improving the IP-to-AS mapping and the technique to compare traceroute paths with AS paths.

\bibliographystyle{plain}
\bibliography{bibliography_extended}

\end{document}

%% file: 020-relatedwork.tex
\section{Related Work}
\label{sec:relatedwork}

Many organizations and projects are devoted to assessing the quality of service offered by ISPs. SamKnows~\cite{samknows} is a free broadband measurement service for consumers. The Federal Communications Commission and SamKnows have joined together to provide American consumers with statistics on their broadband connections. MisuraInternet~\cite{agcom} is an Italian project for measuring the quality of broadband access, supported by the Italian Authority for Telecommunications. Project BISmark~\cite{bismark} is a platform for performing measurements of ISP performance. Related projects such as RIPE Atlas~\cite{atlas}, CAIDA Ark~\cite{ark}, and M-Lab~\cite{mlab} perform large scale active measurements towards several targets.

A framework to analyze the impact of routing changes on network delays between end hosts is presented in~\cite{pzmh-undcre-07}. The authors focus on understanding how the network
delay and jitter measured in a stable routing state change after the routing has switched to a new stable state.
In their main experimental setup they trigger active probing when real-time monitored BGP updates are received.
Their findings indicate that there exists stability in the resulting delay differences for the path before and after the routing event, and correlation between routing changes and delay. Hence, their results strongly motivate our study.
In~\cite{zmz-edrdes-08} the authors diagnose causes for routing events associated with large ISPs, using continuous probing from multiple end systems and building a greedy algorithm to explain multiple events occurring close in time with small sets of causes. They validate their methodology using routing disruptions publicly reported by network operators.
The analysis of end-to-end measurements presented in \cite{5542733} proves that most delay variations are either link or router related, while congestion can be observed rarely.
A tool for the visualization of the correlation between BGP and RTT data is described in~\cite{dds-vdcbbrrtdam-13}; however the correlation is an input of the tool, rather than an outcome.

Many other researchers have used statistical tools for the analysis of trends in network data. In~\cite{msgswyze-dpiulon-10} the authors describe a novel infrastructure for detecting the impact of network upgrades on
performance. Their system extracts interesting so-called ``triggers'' from a large number
of network maintenance activities. It then identifies behavior changes in network
performance caused by the triggers. It uses statistical rule mining and network
configuration to identify commonality across the behavior changes. \cite{mgwyzehs-rdmncsp-11} presents a new tool for detecting
maintenance-induced performance changes in a timely fashion. It uses association between
maintenance and network elements to identify performance metrics for time-series analysis.
It uses a new Multiscale Robust Local Subspace algorithm (MRLS) to accurately identify
changes in performance.

%% file: 025-scenario.tex
\section{Reference Scenario}
\label{sec:reference-scenario}
\begin{figure}
   \centering
   \includegraphics[width=\columnwidth]{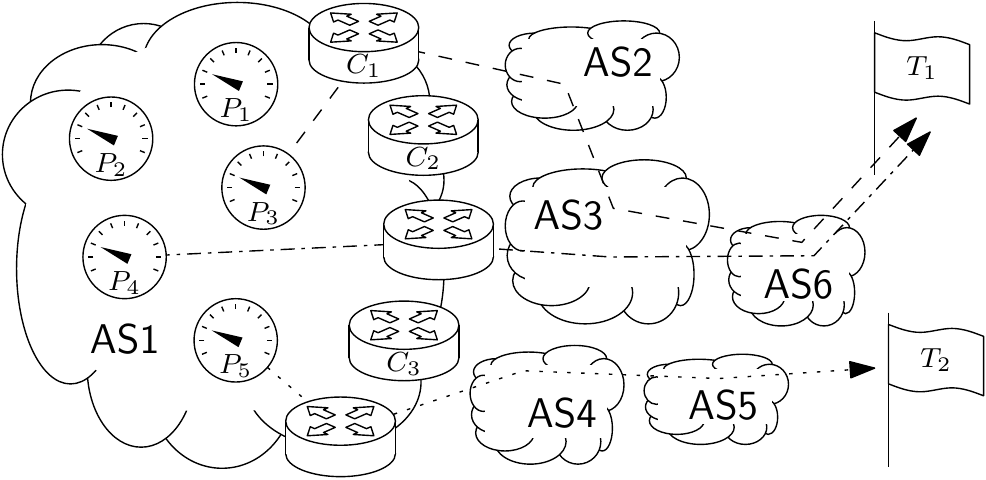}
   \caption{Our reference scenario. $C_i$, $P_i$, and $T_i$ are BGP routers gathering routing information (collector peers), network probes, and targets (IP addresses), respectively. Clouds are Autonomous Systems. Dashed polylines are traffic paths.}
   \label{fig:reference-scenario}
\end{figure}

In order to obtain useful results with our methodology, BGP route collectors and network probes that perform the measurements should be arranged in a consistent way. In particular, our reference scenario for the rest of the paper is illustrated in Fig.~\ref{fig:reference-scenario}.
Each cloud represents a different Autonomous System (AS); we focus on AS1, namely we are interested in matching BGP routing information and network measurements that are gathered within this AS. We assume that the following collection points are available inside AS1: a set of BGP routers $C_i$, called \emph{Collector Peers} (CPs), that forward all the received BGP updates to a central BGP collector for storage, and a set of boxes $P_i$, called \emph{probes}, that perform active measurements on a periodical basis and store them to some other repository. Note that a BGP router is not necessarily a collector peer (see the unnamed routers in Fig.~\ref{fig:reference-scenario}), and a collector peer is not necessarily an egress point for traffic exiting AS1 (as it is the case for $C_2$). A \emph{measurement} is the action (ping, traceroute) that a probe undertakes in order to collect some information (RTT, router-level path) about a forwarding path towards a certain destination, called \emph{target}. In the scenario in Fig.~\ref{fig:reference-scenario}, each probe $P_i$ can perform measurements towards any of the targets $T_i$: the forward path taken by probe traffic, indicated with dashed lines in the figure, may vary depending on the control plane information at intermediate routers and on the target $T_i$. In the specific case when such traffic traverses a collector peer, as it is the case for traffic from $P_3$ to $T_1$, a matching between BGP routing information and network measurements is obviously found. However, even when this does not happen (for example, for $C_2$ and $C_3$), a correlation between BGP routing changes and variations in measurement results may still exist, provided that BGP routing information is available for a BGP prefix that comprises the measurement target under investigation: this is exactly what our methodology helps to discover. To sum up, in the described scenario the possibility to determine this correlation depends on two factors:
\begin{inparaenum}[i)]
   \item the availability of a set of BGP collector peers and network probes within the same AS, and
   \item the availability of BGP routing information for a prefix that comprises the IP address of the target under investigation.
\end{inparaenum}

%% file: 030-methodology.tex
\section{Matching BGP Routing Changes and RTT Variations}
\label{sec:methodology}
We now describe our methodology to automatically determine a correspondence between BGP routing changes and delay variations. Inputs to this methodology consist of data from the control plane of BGP routers and from active RTT measurements performed by network probes.
We highlight that no knowledge is assumed about the network topology, the distribution of the probes, and the distribution of CPs within the ISP.
As a primary output, by using this methodology it is possible to determine whether a specific BGP routing change has an observable effect in the form of a variation in RTT values recorded by a probe.
Our methodology has several tunable parameters: in the following description we assume that they are all set to fixed values. We show in Section~\ref{sec:experiments} how our methodology can be instantiated with different parameter values in order to compute aggregate correlation estimates.

\begin{figure*}
   \centering
   \subfigure[]{\includegraphics[scale=.8]{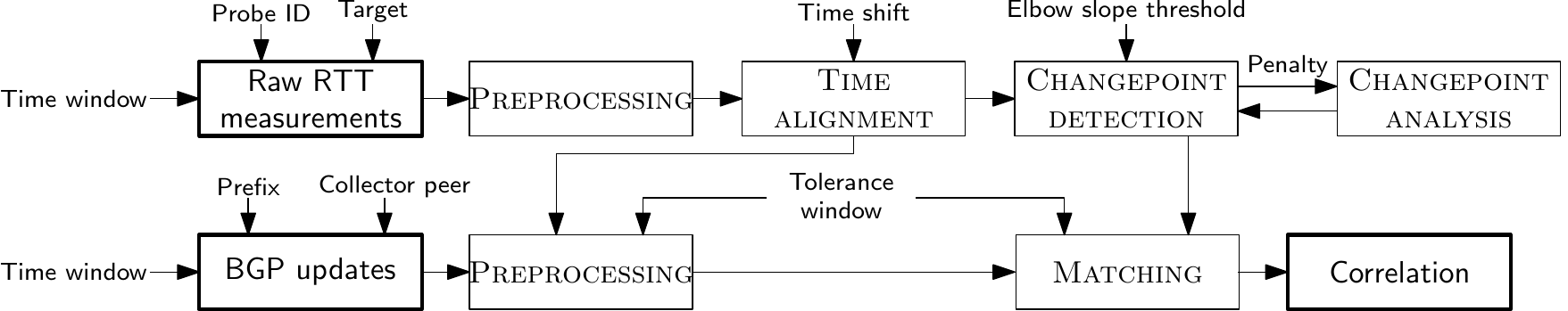}}
   \subfigure[]{\includegraphics[scale=.8]{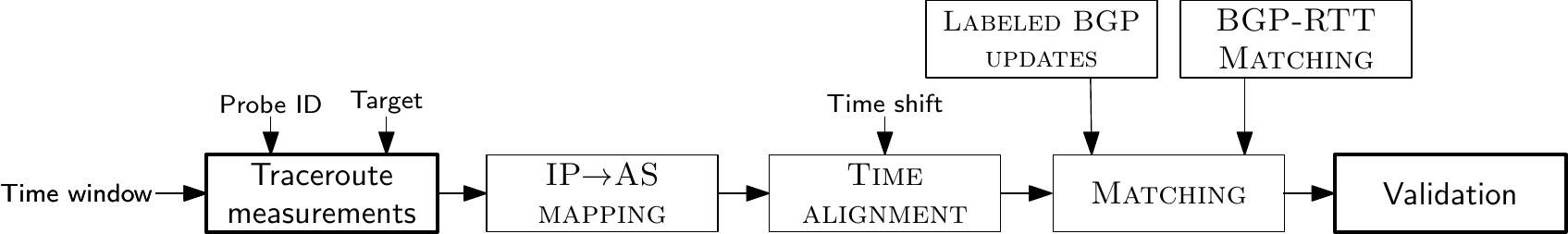}}
   \caption{The main processing steps of our 
   methodology (a) and of its validation (b). Thick-border boxes are 
   input data or output results. Thin-border boxes are operations. 
   Arrows indicate inputs and outputs for each operation. Additional 
   inputs and tunable parameters are represented without a box (all
   occurrences of equally named parameters have the same value).}
   \label{fig:methodology}
\end{figure*}

The main processing steps are represented in\linebreak Fig.~\ref{fig:methodology}(a), which can be used as a reading key for the rest of the section. 
We highlight that both BGP information and measurement information undergo several processing operations before applying the \textsc{Matching} step. In fact, one of the most relevant challenges in matching measurements with BGP routing changes is the intrinsically different nature of the two data sets. Being active measurements, RTTs are usually recorded by probes on a periodical basis, whereas BGP updates are collected at the exact time instants at which they occur: as an immediate consequence, RTT values are available at a constant rate over time, whereas it is common to observe bursts consisting of lots of BGP changes in a short time followed by long periods of stability.
Moreover, RTTs are subject to very frequent variations over time, whereas BGP routing changes involve a limited number of different paths. Given the possibility that the clocks of BGP routers and probes are not perfectly synchronized, we should also take into account the presence of a time difference between the two kinds of data.

\subsection{Processing of RTT Measurements}
\label{sec:rtt-meth}

We start by describing the sequence of processing steps that are applied to RTT measurements.
We consider as input to these steps a sequence of RTT measurements performed on a periodical basis by a probe identified by a \textsf{Probe ID} towards a fixed destination IP address which we call \textsf{Target}. We assume that each RTT measurement consists of a fixed-length sequence of RTT values (as it is often the case, since the measurement is performed using the \texttt{ping} command), augmented with the time instant at which the measurement was performed and with the actual IP address reached by the measurement. Only measurements within a specified \textsf{Time window} are considered for the analysis.

In the \textsc{Preprocessing} step we perform some filtering operations to remove spurious RTT measurements that do not convey useful data. As the sole exception in our methodology, the actions undertaken at this step are tailored to the specific measurement infrastructure we used in our experimentation, namely the RIPE Atlas~\cite{atlas} platform.
Given the format of the data from~\cite{atlas}, RTT measurements are expected to consist of 3 RTT values, of which we only preserve the minimum one. With this choice we aim at focusing on the propagation and transmission delays, whose values are mostly dependent on the length of forward and reverse router paths and on the physical distance of traversed devices; on the other hand, processing and queueing delays are subject to high variability and may occasionally affect individual RTT values (see, e.g.,~\cite{hm-owdmc-07}). We consider measurements that have less than 3 recorded RTT values or that reach an unexpected IP address to be abnormal and we discard them.

We then perform a \textsc{Time alignment} step to account for several important aspects. First of all, the observations of a BGP routing change and of a corresponding variation in the RTT values may be separated by a time lapse. In the general case, we expect the RTT variation to occur after the routing change that caused it. However, this is strongly influenced by the relative position of the CPs and of the probes, because the time required for a change to be propagated may vary according to network conditions or to delays that are artificially introduced by router settings such as the MRAI: depending on these delays, the BGP routing change may have a more recent timestamp than the corresponding RTT variation.
Further, there may be an offset between the clock of the CPs and the clock of the probes.
In order to consider these aspects, we shift the timestamps of RTT measurements by a fixed \textsf{Time shift} value. We will show in Section~\ref{sec:experiments} how to find a suitable value for this parameter.

As discussed above, RTT measurements consist of a sequence of highly variable RTT values. Indeed, it is unlikely to find two consecutive RTT measurements with the same RTT value, and every RTT measurement could be considered as representative of a change in the network delays. The choice of retaining only the minimum value for each RTT measurement in the \textsc{Preprocessing} step can only partially mitigate this issue. As an extreme consequence, almost any BGP routing change could be matched with an arbitrarily chosen RTT measurement that is close enough in time. 
%
To overcome this problem, we introduced a \textsc{Changepoint detection} step. In this step we process RTT measurements in order to find significant changes in their values. We adopt a methodology called \emph{changepoint analysis}, commonly used in statistical data science to evaluate any time series and detect changes in the data (for a survey, see~\cite{bn-dacta-93}). Given the nature of RTT values, we are interested in detecting the time instants at which the mean value changes persistently. We use a recent technique called PELT~\cite{kfe-odcwlcc-12}, which is based on an exact and efficient algorithm to detect both mean and variance shifts in time series data. PELT requires an input parameter called ``penalty'' that affects the precision of the analysis: a low value means that the algorithm should consider also volatile mean shifts as valid changes, whereas a high value only takes into account mean shifts that span a considerable portion of the whole input. The effect of different choices of the penalty on the number of detected changepoints is depicted in Fig.~\ref{fig:penalty-changepoint-hyperbole}.
We chose PELT because it is a very efficient variation of a technique called Optimal Partitioning~\cite{1381461}, which is considered the state of the art in the field. To apply the technique, we relied on a publicly available implementation in R~\cite{r-changepoint-package}.

Choosing the right penalty value is not a simple task. We adopt a rule called \emph{elbow method} (see, e.g.~\cite{SMJ:SMJ819}, where the method is applied to a clustering problem), traditionally used by statisticians on datasets that show the same distribution as in Fig.~\ref{fig:penalty-changepoint-hyperbole}: the chosen penalty is the one for which the number of changepoints starts decreasing at a slow pace for increasing values of the penalty. The underlying intuition is that the PELT algorithm first identifies the largest changes and then starts to add smaller ones, eventually considering noise from the data as legitimate mean variations. Hence, given an \textsf{Elbow slope threshold} that represents an acceptable difference quotient between the variation of penalty and the corresponding variation in the number of changepoints, we enter an iterative process whose building block is called \textsc{Changepoint analysis}. We start with a low penalty $p_{0}$ and compute the number of changepoints $cpt_{0}$. Then for $i = 1, \dots, n$ we compute a new penalty $p_{i} = f(i)$ (with $f(0)=p_0$), where $f$ is a strictly increasing function, and the number of changepoints $cpt_{i}$. The loop ends for $i = i^{*}$ such that the difference quotient $\delta = \frac{p_{i^{*}} - p_{i^{*} -1}}{cpt_{i^{*}-1} - cpt_{i^{*}}}$ is less than the specified \textsf{Elbow slope threshold}. At this point we use the penalty $p_{i^{*}}$ to transform the original sequence of RTT measurements into a time-labeled step function.

On a side note, the \textsc{Changepoint analysis} requires a simple sequence of samples as input, without taking into account their timestamps. When a candidate sample is elected as changepoint we associate the change with the original timestamp of the sample. That means we are potentially introducing an error of magnitude equal to the period of RTT measurements. In Section~\ref{sec:correlation-meth} we explain how to mitigate such issue when matching BGP routing changes with RTT changepoints.

\subsection{Processing of BGP Routing Changes}
\label{sec:bgp-meth}

The analysis of BGP information also involves a number of steps before the matching with RTT measurements. The input data consists of a sequence of routing changes observed by a \textsf{Collector peer} in the BGP path used to reach a specified \textsf{Prefix}. Each routing change contains the timestamp at which the change was observed and the new AS-path (i.e. a sequence of AS numbers), or an empty sequence when the CP observes a route withdrawal. Similarly to what happens with RTT measurements, we only consider routing changes recorded within a specified \textsf{Time window}.

The \textsc{Preprocessing} step is needed to determine which BGP routing changes are eligible for further analysis. This filtering procedure is based on the outcome of the \textsc{Time alignment} step described in Section~\ref{sec:rtt-meth}. While in principle every single BGP event is crucial, our filtering is motivated by the nature of the RTT measurements. First of all we need to take into account the rate at which RTT measurements are collected (one every 4 minutes for RIPE Atlas~\cite{atlas}): for each pair of consecutive RTT measurements we ignore all BGP routing changes happening in between, except the last one. Further, we simply ignore all BGP routing changes occurring between every pair of consecutive RTT measurements that are separated by a time lapse greater than an input \textsf{Tolerance window} (which should be larger than the time between two consecutive RTT measurements). Both decisions have the effect of isolating BGP routing changes that can not be ``seen'' in the available RTT measurements, therefore avoiding any improper deductions on them.

\subsection{Matching and Correlation}
\label{sec:correlation-meth}

Once both the RTT measurements and BGP routing changes are cleaned up and processed, we can look for a correspondence between the two datasets. 

The role of the \textsc{Matching} step is to determine whether each BGP routing change corresponds to one or more RTT changepoints. For each preprocessed BGP routing change with timestamp $t$, we center the \textsf{Tolerance window} (described in Section~\ref{sec:bgp-meth}) at $t$ obtaining a \emph{matching window}. We use the latter to filter RTT changepoints by timestamp and associate those falling within a matching window with the BGP routing change where that window is centered. Each BGP routing change is then marked as ``correlated'' if there exists at least one RTT changepoint associated with it.

Finally, to produce the \textsf{Correlation} estimate, we divide the number of correlated BGP routing changes by the total number of preprocessed BGP routing changes, obtaining a normalized value that we call \emph{BGP-RTT correlation factor}.

%% file: 050-experiments.tex
\section{Experiments and Validation}
\label{sec:experiments}

In this section we describe the data sets we considered to validate our methodology and discuss the results of our experiments. We then further validate these results by using of traceroute data.

\subsection{Data Sets}
\label{sec:experiments-datasets}

The first condition to apply our methodology is the availability of data sets matching the reference scenario in Section~\ref{sec:reference-scenario}. Traditionally, researchers analyzed BGP and RTT information by exploiting the data sources mentioned in Section~\ref{sec:relatedwork} or by performing ad-hoc measurements to build up a customized data set (e.g.,~\cite{pzmh-undcre-07}).
%
In this paper we make use of data collected within two projects of growing popularity: the RIPE Routing Information Service~\cite{ris}, that offers hundreds of worldwide spread CPs, and RIPE Atlas~\cite{atlas}, that gathers data from thousands of probes performing pings, traceroutes, and other measurements. Besides making notoriously rich data sources available to the public, both projects are maintained by the same independent and non-profit organization, which increases the probability of finding ASes for which there are both CPs and network probes and virtually eliminates from the data any biases deriving from the interests of a specific ISP.

As a preliminary step, we determined a set of ASes suitable for our analysis: we found out that in the RIS and Atlas data sets there were 55 ASes that had at least one active CP and one active probe as of January 2013. All these ASes host a total of 126 CPs (number of unique CPs seen by RIS collectors between January 2011 and December 2012) and 200 probes. In Table~\ref{characterization-as-probe-cp} we show a more precise characterization of the set of selected ASes terms of number of available CPs and probes. Although most of the ASes only match the minimum requirement (i.e., one collector peer and one probe), there are many interesting exceptions.
We remark that Atlas probes are configured with both an IPv4 AS number and an IPv6 AS number, which may in general be distinct: in performing the selection of ASes, we only considered the IPv4 AS number.

\begin{table}
   \centering
   \begin{tabular}{|*{10}{c|}}
      \cline{3-10}
      \multicolumn{2}{c}{} & \multicolumn{8}{|c|}{\footnotesize\textbf{Number of probes}}\\
      \cline{3-10}
      \multicolumn{2}{c|}{} & \textbf{1} & \textbf{2} & \textbf{3} & \textbf{4} & \textbf{5} & \textbf{7} & \textbf{13} & \textbf{22}  \\
      \hline
      \multirow{8}{*}{\rotatebox{90}{\footnotesize\textbf{Number of CPs}}} &
      \textbf{1} & 22 & & 1 & 1 & & & & \\
      \cline{2-10}
      & \textbf{2} & 12 & 3 & 3 & 2 & & & & 1 \\
      \cline{2-10}
      & \textbf{3} & 1 & & & 1 & 1 & & & \\
      \cline{2-10}
      & \textbf{4} & 1 & 1 & & & & & & \\
      \cline{2-10}
      & \textbf{5} & 1 & & & & & & 1 & \\
      \cline{2-10}
      & \textbf{6} & 1 & & & & & 1 & & \\
      \cline{2-10}
      & \textbf{7} & & & 1 & & & & & \\
      \cline{2-10}
      \hline
   \end{tabular}
   \caption{Available probes and CPs in the data sets we used. Each cell contains a count of the ASes for which the corresponding number of probes and CPs are available. Empty cells are equivalent to a zero count.}
   \label{characterization-as-probe-cp}
\end{table}

We then selected a set of interesting targets. Within the Atlas infrastructure, probes periodically perform measurements (typically pings, traceroutes, and HTTP requests) towards a set of predefined targets that include all the root name servers and some Web services exposed by the RIPE NCC: we selected the most interesting targets according to the following criterion. For each target we used the RIPEstat service~\cite{ripestat} to determine the most specific IP prefix containing it and seen by RIS route collectors, and we analyzed the BGP and RTT data available for that target. Considering the fact that RIPE Atlas has only been active since late 2010, we extended the analysis over a time window of two years, thus fixing the \textsf{Time window} parameter in Fig.~\ref{fig:methodology} for all subsequent experiments from January 2011 until December 2012. We downloaded the following data for all the 23 available targets and corresponding IP prefixes: 
\begin{inparaenum}[(a)]
	\item BGP updates and table dumps collected for all IP prefixes by all available RIS collector peers;
	\item RTT measurements (performed every 4 minutes) and traceroute measurements (performed every 20 minutes) collected by all the Atlas probes.
\end{inparaenum}
We then estimated the total amount of available data for each target and IP prefix, in order to restrict our study to a more meaningful subset of targets. Table~\ref{target-data-size-tradeoff} contains the targets that were selected, because they have a good trade-off between the amount of available BGP routing changes and RTT measurements, because they contain a sample of both anycast and unicast prefixes, and because they are associated with IP prefixes of different lengths.
\begin{table*}
   \centering
   \begin{tabular}{|c|l|ll|r|r|}
      \cline{1-4}
      \multicolumn{4}{|c|}{Target}\\
      \hline
      ID & IP address & \multicolumn{2}{l|}{\parbox{2.5cm}{IP prefix\\(Unicast/Anycast)}} & \parbox{2cm}{RTT data\\(MBytes)} & \parbox{2cm}{BGP data\\(updates)}\\
      \hline
      1001 & 193.0.14.129 (\texttt{k.root-servers.net}) & 193.0.14.0/24 & (A) & 16,168 & 97,871 \\
      \hline
      1003 & 193.0.0.193  (\texttt{ns.ripe.net}) & 193.0.0.0/21 & (U) & 15,953 & 9,541 \\
      \hline
      1004 & 192.5.5.241  (\texttt{f.root-servers.net}) & 192.5.5.0/24 & (A) & 16,005 & 35,553 \\
      \hline
      1005 & 192.36.148.17 (\texttt{i.root-servers.net}) & 192.36.148.0/24 & (A) & 16,144 & 13,435 \\
      \hline
   \end{tabular}
   \caption{Amount of available data for the selected ASes and measurement targets.}
   \label{target-data-size-tradeoff}
\end{table*}

\subsection{Parameter Tuning and Correlations}\label{sec:tuning}
\begin{figure}
   \centering
   \includegraphics[width=.8\columnwidth]{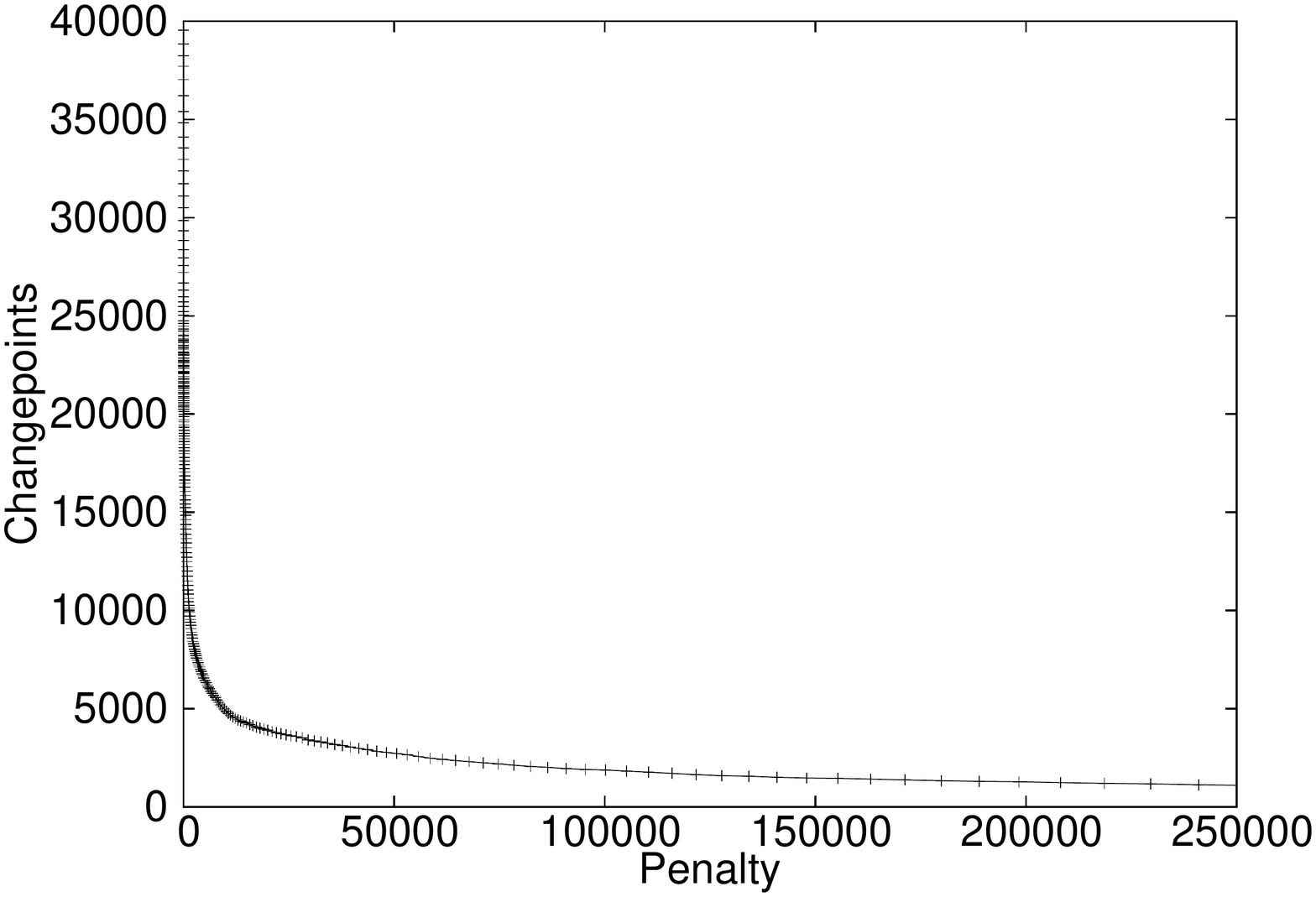}
   \caption{Relationship between the penalty value and the number of changepoints found by the PELT algorithm. The plot was produced for measurement 1001, for data from the Atlas probe \#167, and with penalty values defined as $p_i = 1.05^i - 1$, $i\geq 1$.}
   \label{fig:penalty-changepoint-hyperbole}
\end{figure}
We cleaned them up the downloaded data in the respective \textsc{Preprocessing} steps and then determined the correlation between RTT measurements and BGP updates as described in Section~\ref{sec:methodology}. First of all we processed all RTT measurements using the PELT technique, with penalty values selected according to the elbow method. Since at the beginning we had no hints about the impact of the \textsf{Elbow slope threshold} on the penalty and, therefore, on the set of computed changepoints, we executed the PELT implementation once for each of the following 12 \textsf{Elbow slope threshold} values: 0.001, 0.01, 0.1, 1, 5, 10, 50, 100, 200, 300, 1000, 10000. We chose these values because they are representative of a wide range of slopes in the penalty/changepoints curve (see Fig.~\ref{fig:penalty-changepoint-hyperbole}), resulting in different penalty values and, therefore, a different accuracy in searching for changepoints. The function used to determine the penalty values during the \textsc{Changepoint detection} was $p_i = c_1^i + c_2$, namely we considered exponentially increasing penalty values. We made this choice because we experimentally verified that the number of changepoints that are no longer detected by the PELT technique when increasing the penalty decreases as the ratio $p_i/p_{i-1}$ decreases. We performed preliminary computations with different values of $c_1$ and $c_2$ (one is shown in Fig.~\ref{fig:penalty-changepoint-hyperbole}), and found that $c_1=2$ and $c_2=0$ was a good compromise between computational time and granularity of the applied penalty values. We therefore pre-computed a clean set of RTT measurements where only significant RTT value changes were retained.

\begin{figure}
   \centering
   \includegraphics[width=1.1\columnwidth]{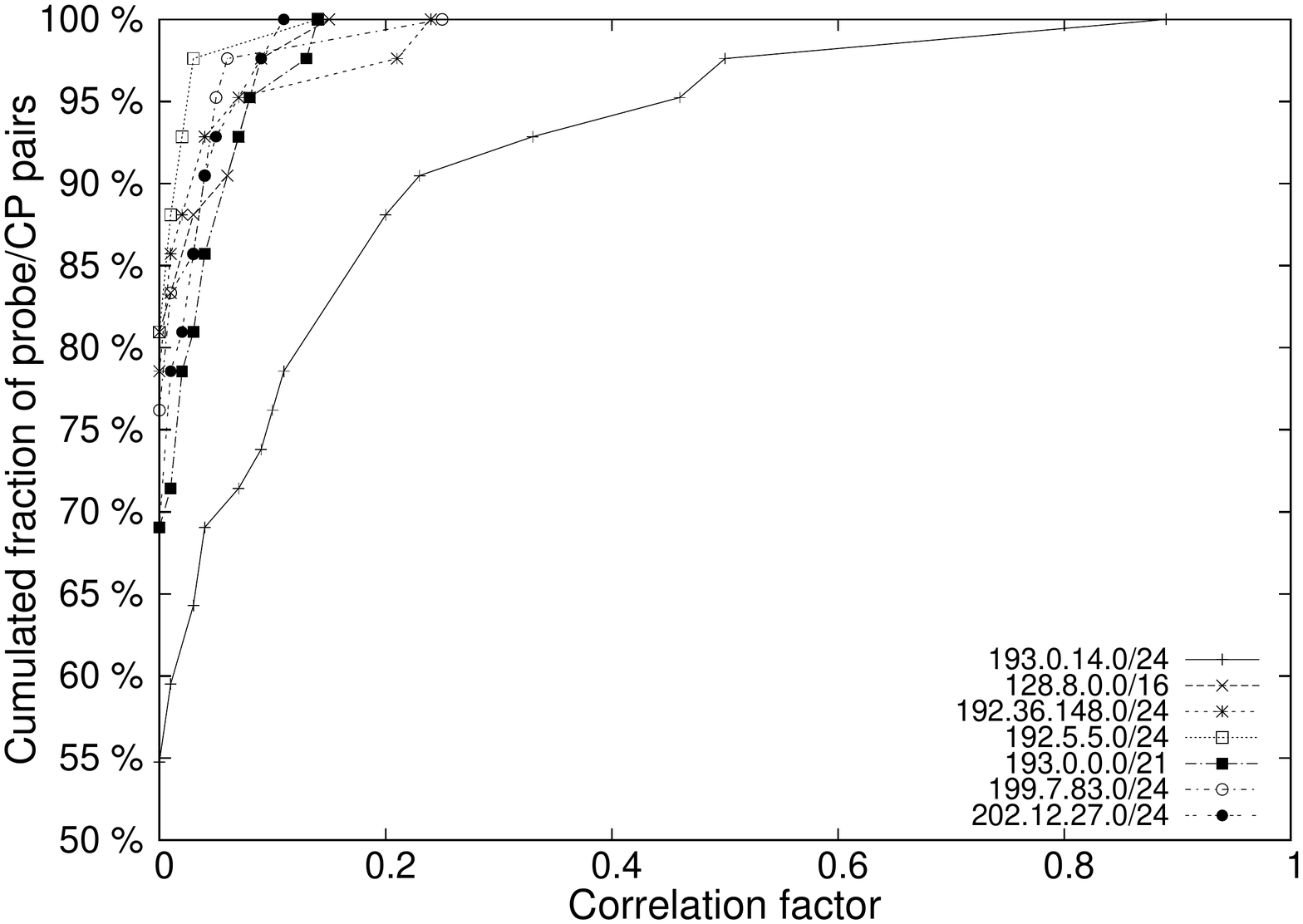}
   \caption{Cumulated fraction of probe/CP pairs exhibiting different values of the correlation factor. Each plot corresponds to a different IP prefix. Only probes that collected RTT data for measurement 1001 and only CPs that recorded BGP updates for all the 7 prefixes are considered. The EST was fixed to 0.001 and the time shift to 0.}
   \label{fig:cdf}
\end{figure}

We then applied the methodology with fixed values of the \textsf{Time shift} and of the \textsf{Elbow slope threshold}, in order to better understand how to distinguish ``good'' correlation values from ``bad'' ones. In particular, we focused on a single \textsf{Target} and considered the RTT values measured towards this target: a set consisting of all the probes that recorded these RTT values was correspondingly identified. We also fixed an IP \textsf{Prefix} and considered the BGP updates for that prefix: a set consisting of all the CPs that recorded these updates was likewise identified. At this point, we composed all the possible pairs consisting of a probe from the first set and a CP from the second set and computed the BGP-RTT correlation factor for each such pair.
We then computed a distribution describing the frequency with which different values of the correlation factor occurred. In order to compare ``good'' values of the correlation factors with ``bad'' values, we repeated the same computations by keeping the \textsf{Target} (and the corresponding probe data) fixed and by picking different choices for the \textsf{Prefix}: for this purpose, we considered a random sample of 7 IP prefixes, selected among those comprising at least one measurement target: 128.8.0.0/16, 192.5.5.0/24, 192.36.148.0/24, 193.0.0.0/21, 193.0.14.0/24, 199.7.83.0/24, and\linebreak 202.12.27.0/24.
We then analyzed the obtained results by plotting for each considered prefix the Cumulative Distribution Function (CDF) of the fraction of probe/CP pairs having different values of the correlation factor. A sample plot is shown in Fig.~\ref{fig:cdf}, where the \textsf{Target} was fixed to 193.0.14.129, (measurement 1001 in Table~\ref{target-data-size-tradeoff}).
In this figure, the X axis represents values of the BGP-RTT correlation factor and the Y axis represents the cumulated fraction of probe/CP pairs for which these values occurred. In order to make correlation factors comparable, in this plot we only considered CPs that recorded BGP updates for all the 7 selected prefixes. Since the \textsf{Target} was fixed to 193.0.14.129, correlation factors computed for \textsf{Prefix} 193.0.14.0/24 were expected to be higher, and in fact the corresponding CDF is shifted to the right. Interestingly, only the CDF for this prefix exhibits this particular shape, while CDFs for all the other prefixes ramp up very quickly and are much similar to each other. We found this feature to recur in all our experiments, and considered it as a distinguishing mark between well-correlated BGP and RTT information and badly correlated information.
Based on this observation, we introduced a more aggregate correlation measure which, instead of considering single probe/CP pairs, characterizes the relationship between a set of RTT measurements for a \textsf{Target} and a set of BGP routing changes for a \textsf{Prefix}. We called this measure \emph{correlation score}, and computed it as the area subtended by the CDF of correlation factors for all the probe/CP pairs corresponding to the \textsf{Target} and \textsf{Prefix} of interest. Note that a lower correlation score indicates a better correlation.
For example, for the fixed \textsf{Target} considered in Fig.~\ref{fig:cdf} the ``good'' correlation score with BGP data for the \textsf{Prefix} that comprises the \textsf{Target} is 0.93, whereas the correlation scores for data for other BGP prefixes range between 0.98 and 0.99.

\begin{figure*}
   \centering
   \begin{tabular}{*{2}{m{6cm}}m{2.9cm}}
      \includegraphics[width=.85\columnwidth]{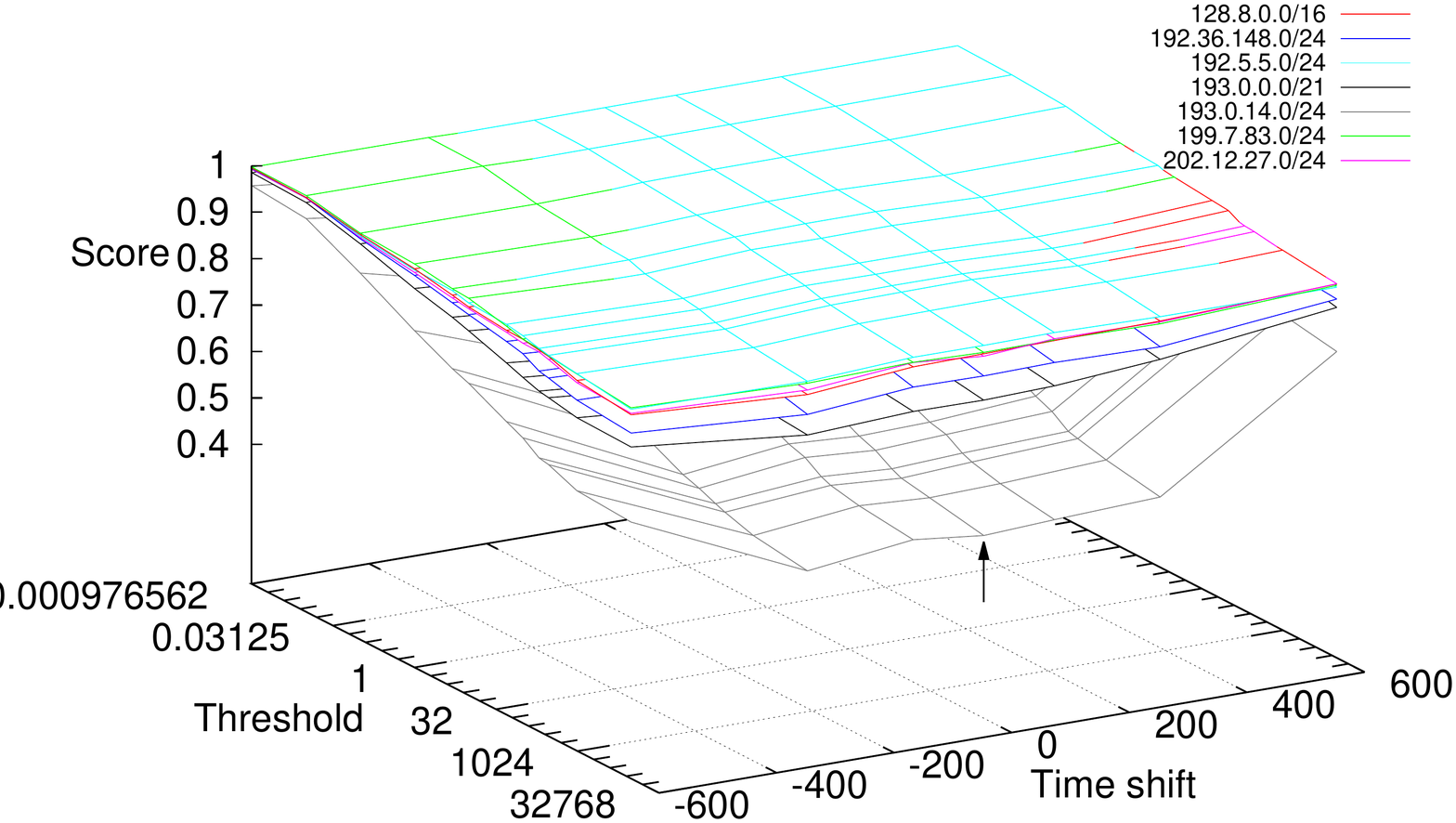} &
      \includegraphics[width=.85\columnwidth]{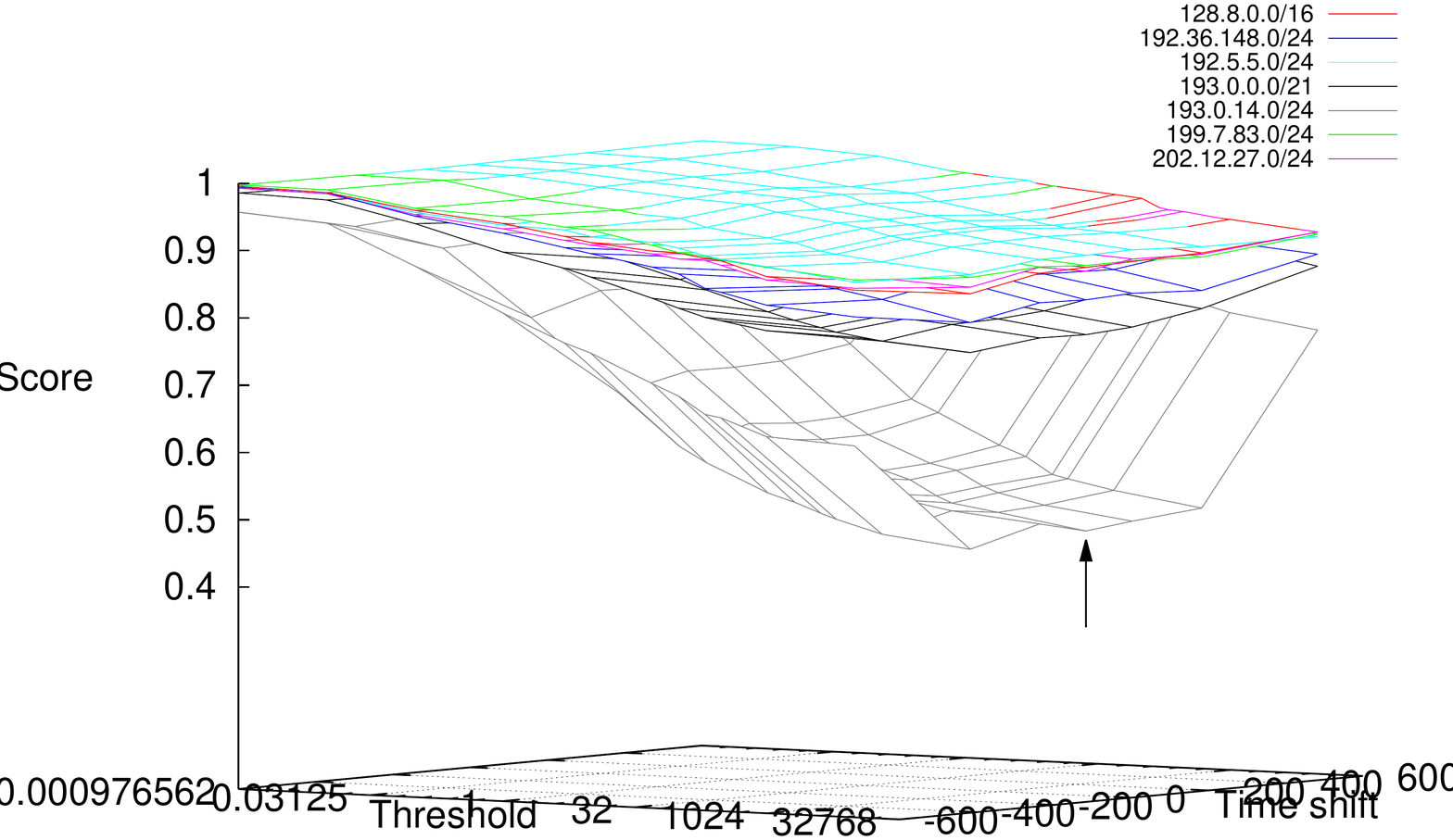} &
      \parbox{3.5cm}{Meas. 1001\\ (target: 193.0.14.129)}\\

      \includegraphics[width=.85\columnwidth]{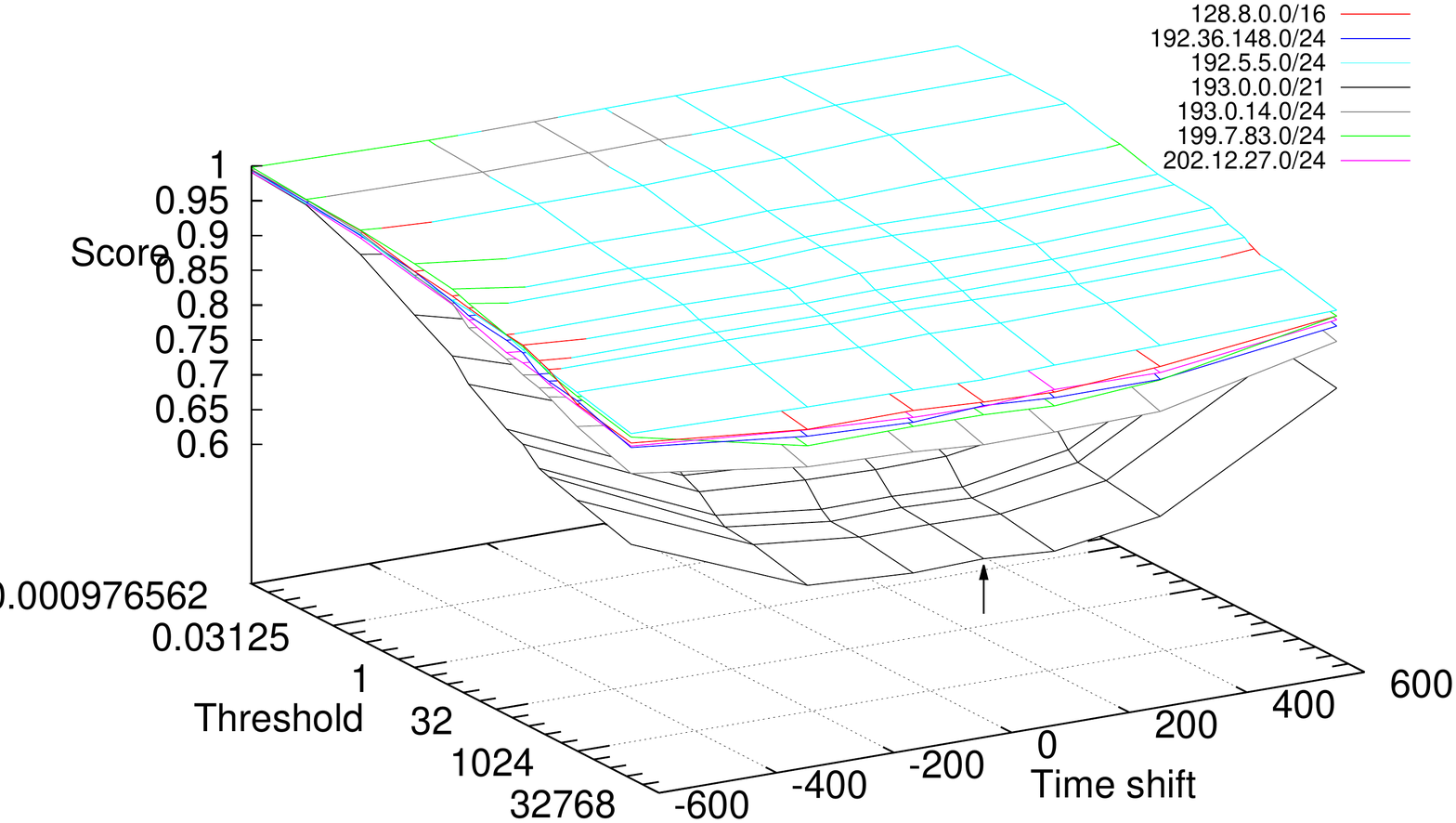} &
      \includegraphics[width=.85\columnwidth]{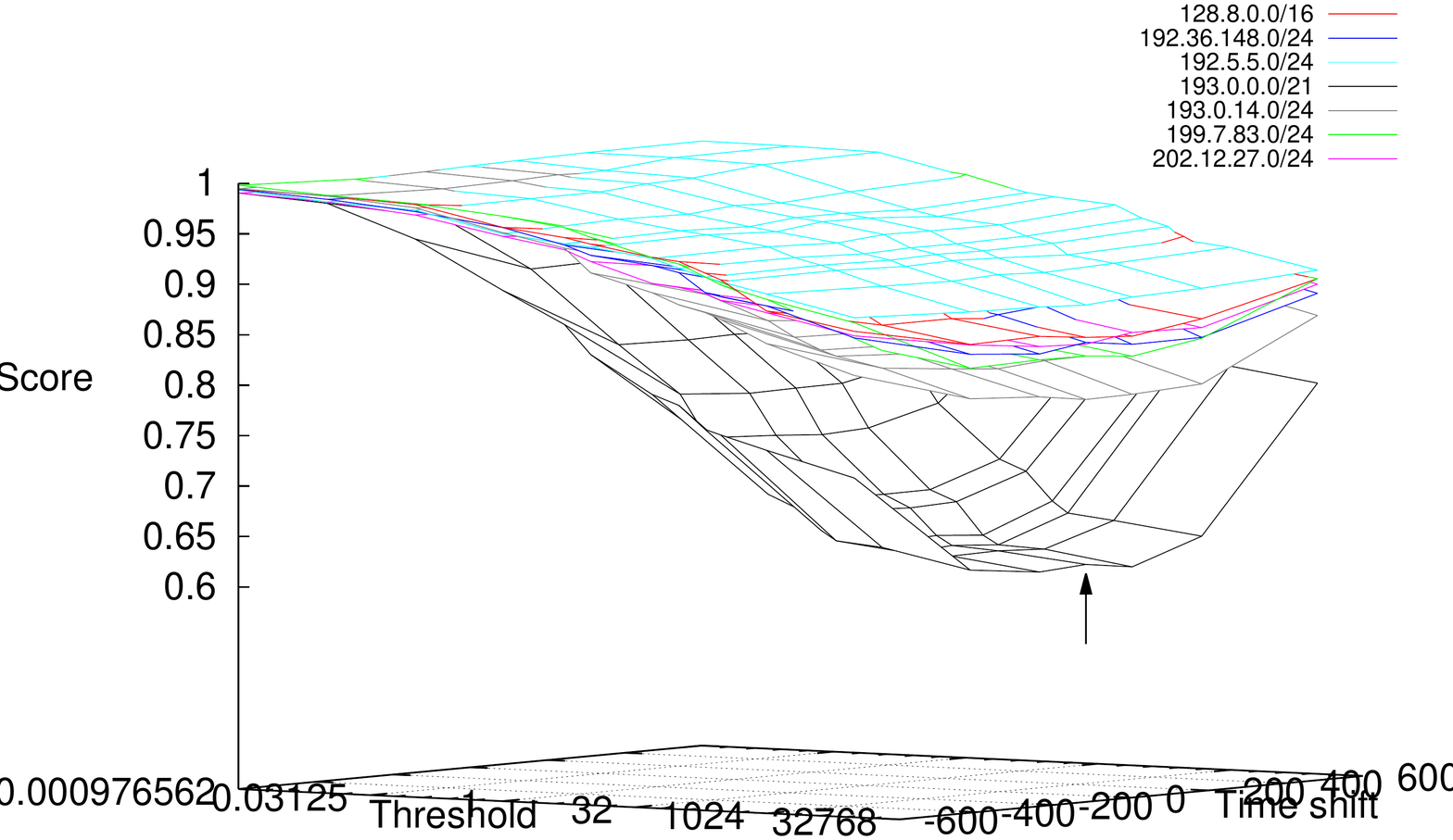} &
      \parbox{3.5cm}{Meas. 1003\\ (target: 193.0.0.193)}\\
      
      \includegraphics[width=.85\columnwidth]{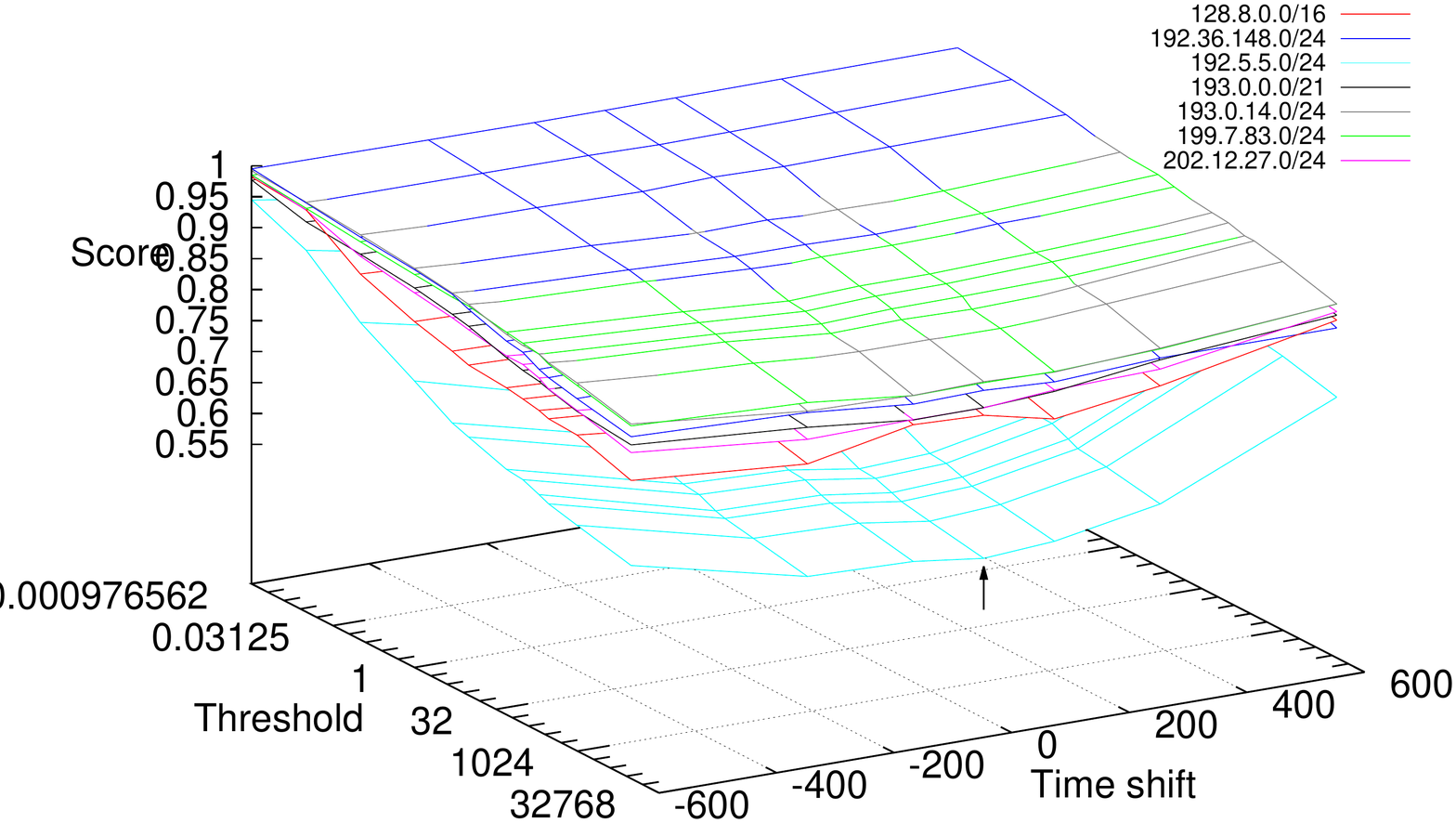} &
      \includegraphics[width=.85\columnwidth]{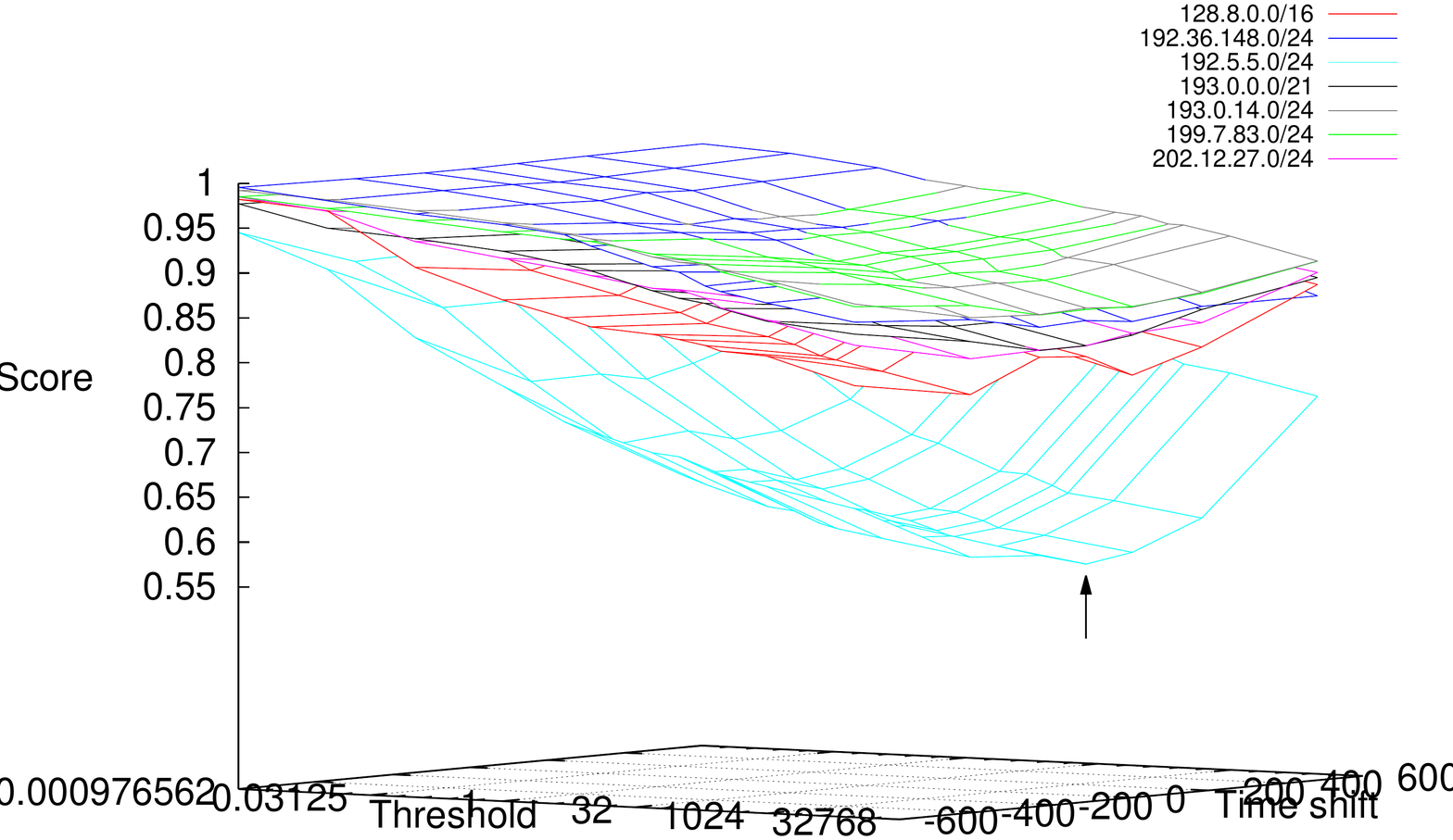} &
      \parbox{3.5cm}{Meas. 1004\\ (target: 192.5.5.241)}\\
      
      \includegraphics[width=.85\columnwidth]{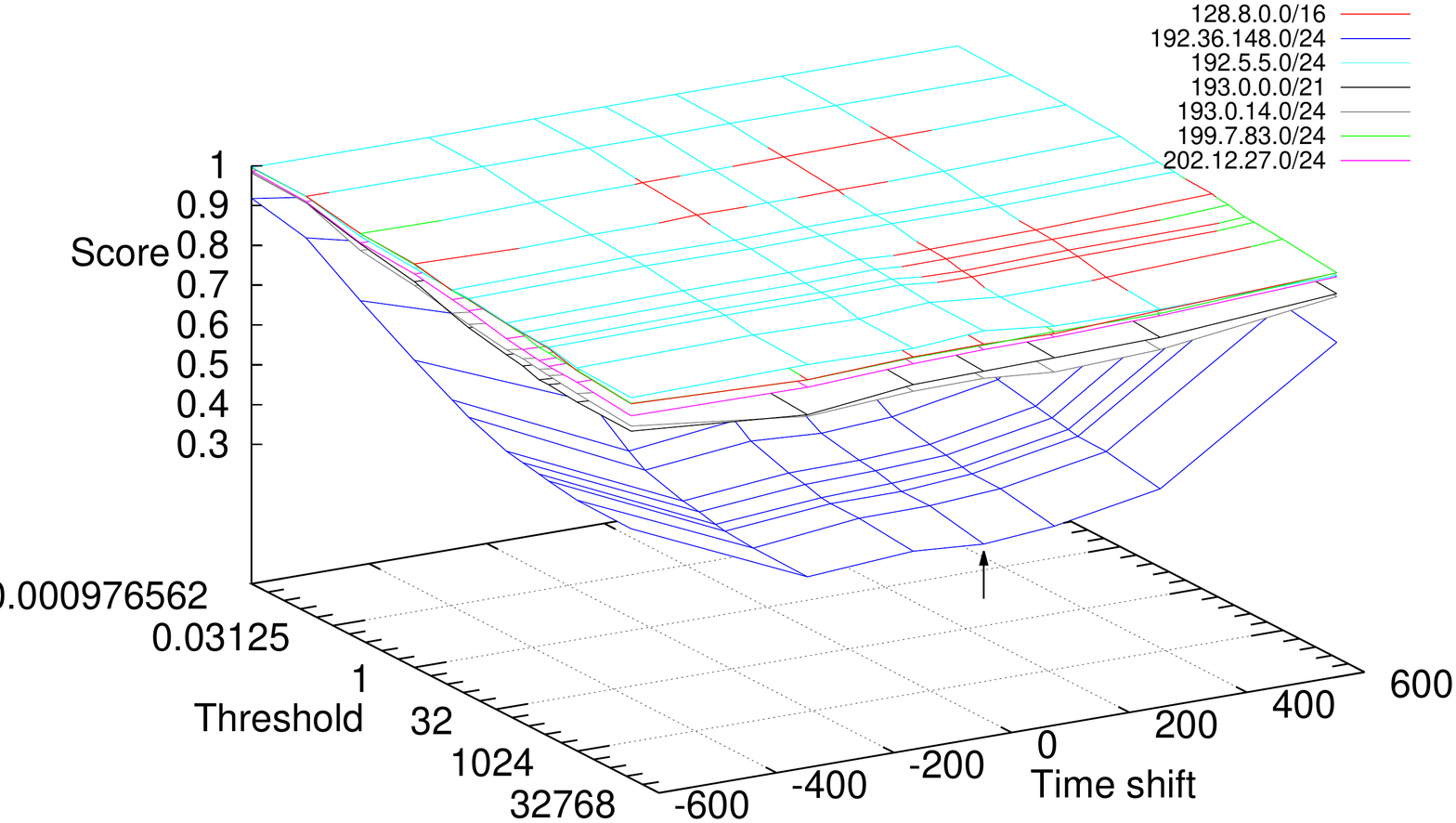} &
      \includegraphics[width=.85\columnwidth]{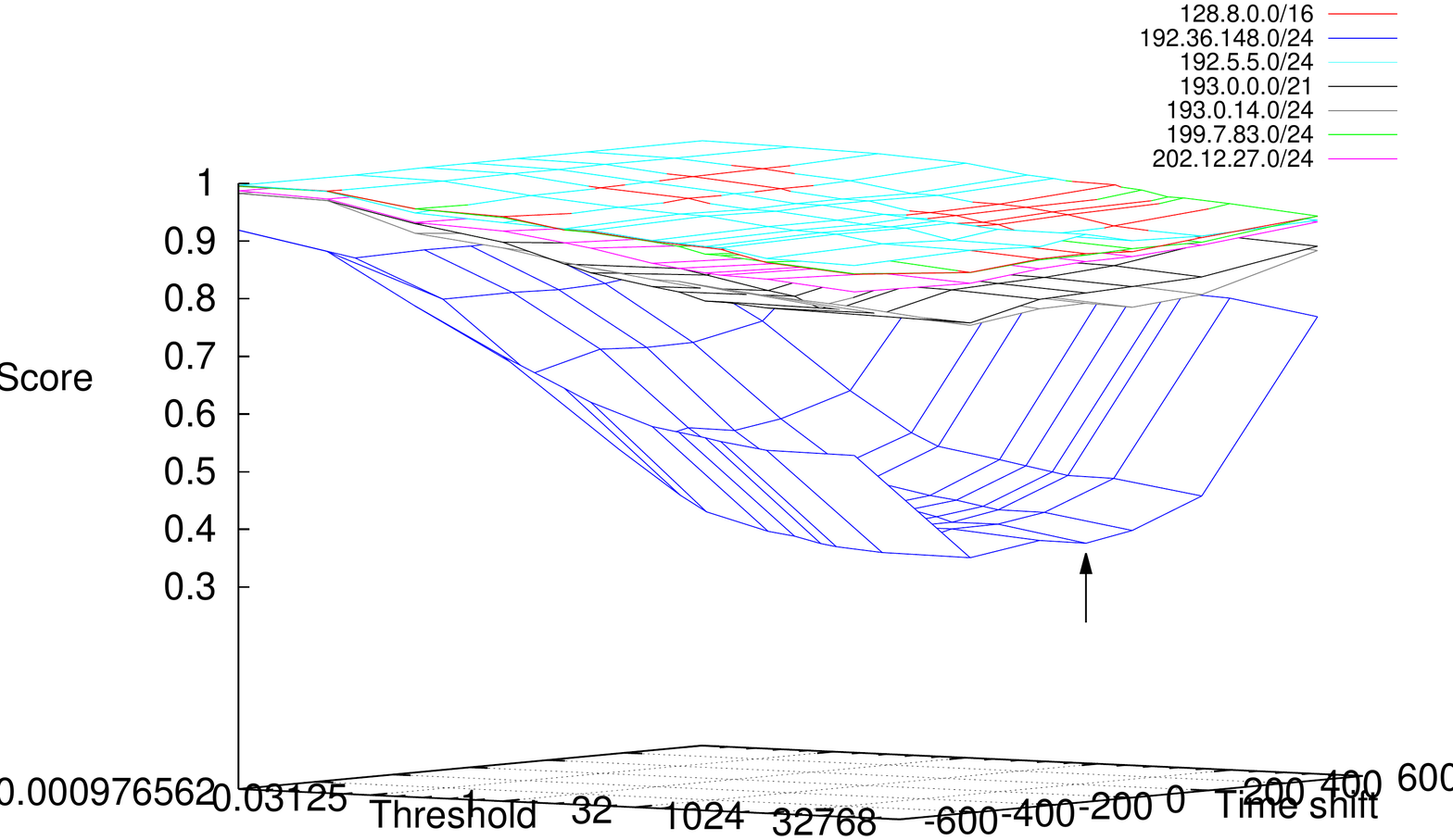} &
      \parbox{3.5cm}{Meas. 1005\\ (target: 192.36.148.17)}\\
   \end{tabular}

   \caption{3D plots showing the value of the BGP-RTT correlation score obtained for different values of the \textsf{Elbow slope threshold} and of the \textsf{Time shift}. Each row contains plots computed for a specific \textsf{Target}, and two views are shown for each plot. Each surface represents correlation scores for a different BGP prefix. The arrow indicates a choice of the \textsf{Elbow slope threshold} and of the \textsf{Time shift} that improves the difference between ``good'' and ``bad'' correlation scores.}
   \label{fig:parameter-choice}
\end{figure*}

After introducing the correlation score, we could more easily assess the influence of the \textsf{Elbow slope threshold} and of the \textsf{Time shift} on the computed correlation values. We considered all the 4 measurements in Table~\ref{target-data-size-tradeoff} and computed the correlation score for each combination of a \textsf{Target} of these measurements and a \textsf{Prefix} among the above selected sample of 7. We repeated the same computation for several combinations of \textsf{Elbow slope threshold} and \textsf{Time shift} values. We picked the first in the set of 12 values, introduced in Section~\ref{sec:tuning}, that we used to pre-compute RTT changepoints, and the second in the following set of values (specified in seconds): $-600$, $-300$, $-120$, $0$, $120$, $300$, $600$. Fig.~\ref{fig:parameter-choice} shows the outcome of this analysis. Plots in each row correspond to a specific \textsf{Target}. Left-side plots and right-side plots are simply two views of the same plot. Each surface refers to a different \textsf{Prefix}. Several conclusions can be drawn by looking at the plots.
First of all, increasing the \textsf{Elbow slope threshold} results in a better distinction between ``good'' and ``bad'' correlation scores: this can be appreciated by observing the separation between the surface corresponding to the \textsf{Prefix} that comprises the \textsf{Target} under consideration (lowest surface in all the plots) and the surfaces corresponding to other \textsf{Prefix}es. The motivation is that higher values of the \textsf{Elbow slope threshold} result in lower penalties used during the \textsc{Changepoint detection} which, in turn, cause more RTT changepoints to be detected by the \textsc{Changepoint analysis}, resulting in higher chances for a successful matching with BGP updates. Moreover, \textsf{Elbow slope threshold}s larger than 10000 make little sense, since unrelated \textsf{Prefix}es start to be improperly correlated with the considered \textsf{Target}: in fact, top-level surfaces start to lower down for higher \textsf{Elbow slope threshold} values (recall that lower scores correspond to better correlation), and this is due to the existence of more RTT changepoints that match BGP updates. In addition, for the \textsf{Prefix} that matches each \textsf{Target}, the correlation score improves significantly for specific values of the \textsf{Time shift}.

From the results conveyed by the plots in Fig.~\ref{fig:parameter-choice} we could determine that picking an \textsf{Elbow slope threshold} equal to 10000 and a \textsf{Time shift} equal to 0, indicated by an arrow in the plots, results in the best separation between ``good'' and ``bad'' correlation scores. Interestingly, these values are extremely close to the optimal choice of parameters for all the measurements we considered, and therefore fixed it throughout the rest of the experiments.
By performing a similar analysis, and by considering the rate of RTT measurements (one every 4 minutes), we also fixed the \textsf{Tolerance window} at 16 minutes.

\subsection{Equivalence Classes}
\begin{figure*}
   \centering
   \subfigure[AS 7575, Measurement 1003 (target: 193.0.0.193)]{%
   \includegraphics[width=\columnwidth]{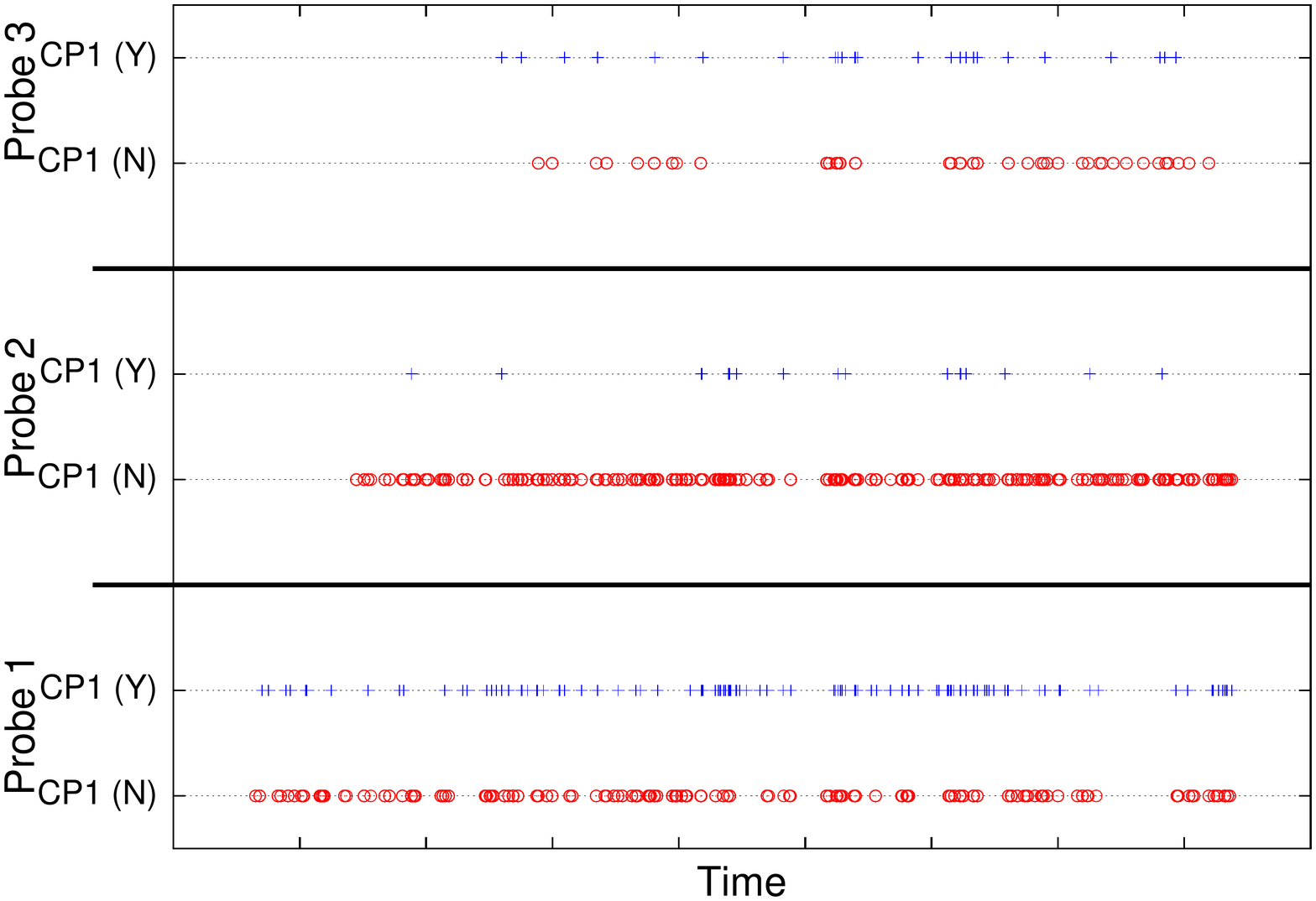}
   \includegraphics[width=\columnwidth]{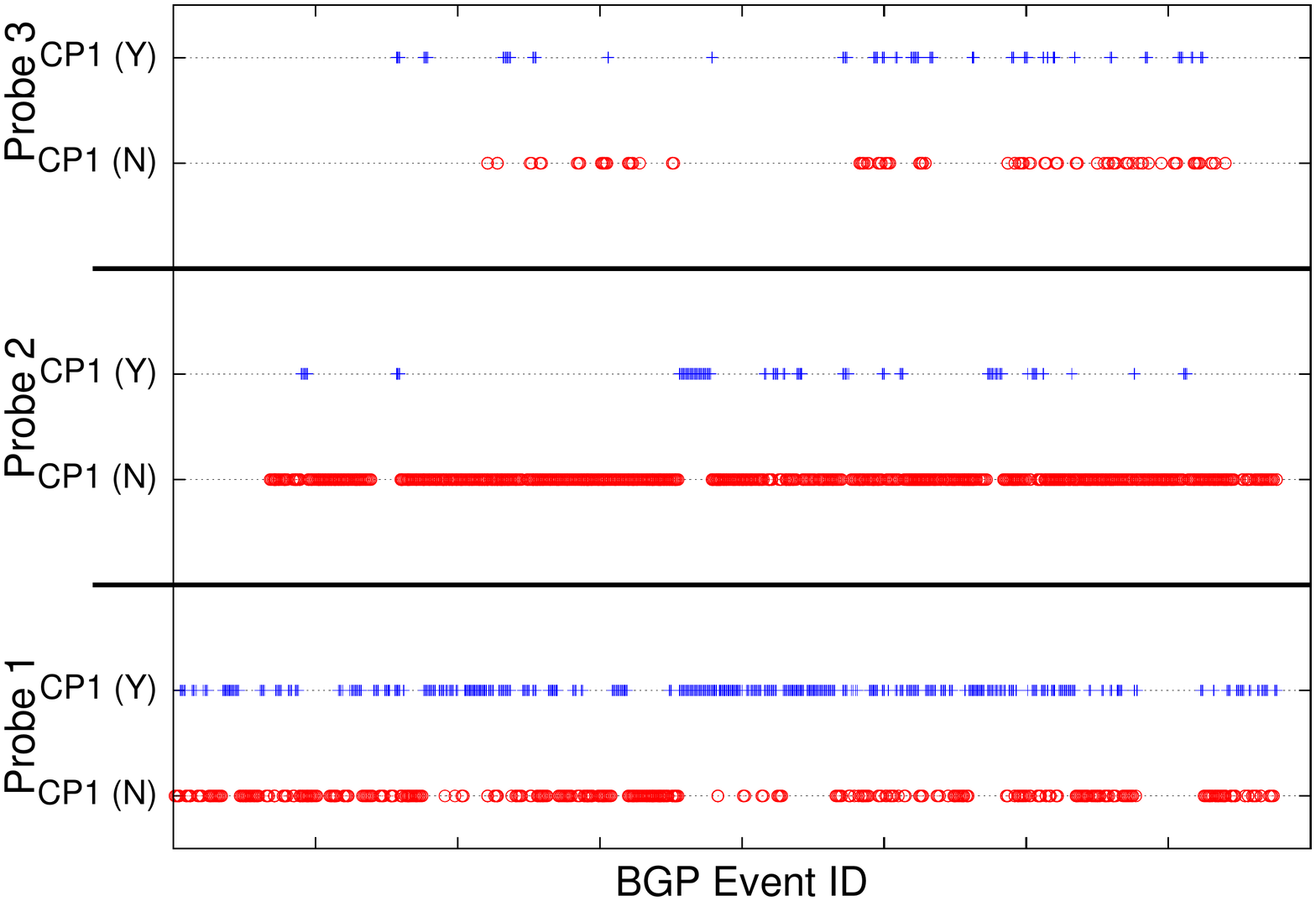}}
   \subfigure[AS 513, Measurement 1004 (target: 192.5.5.241)]{%
   \includegraphics[width=\columnwidth]{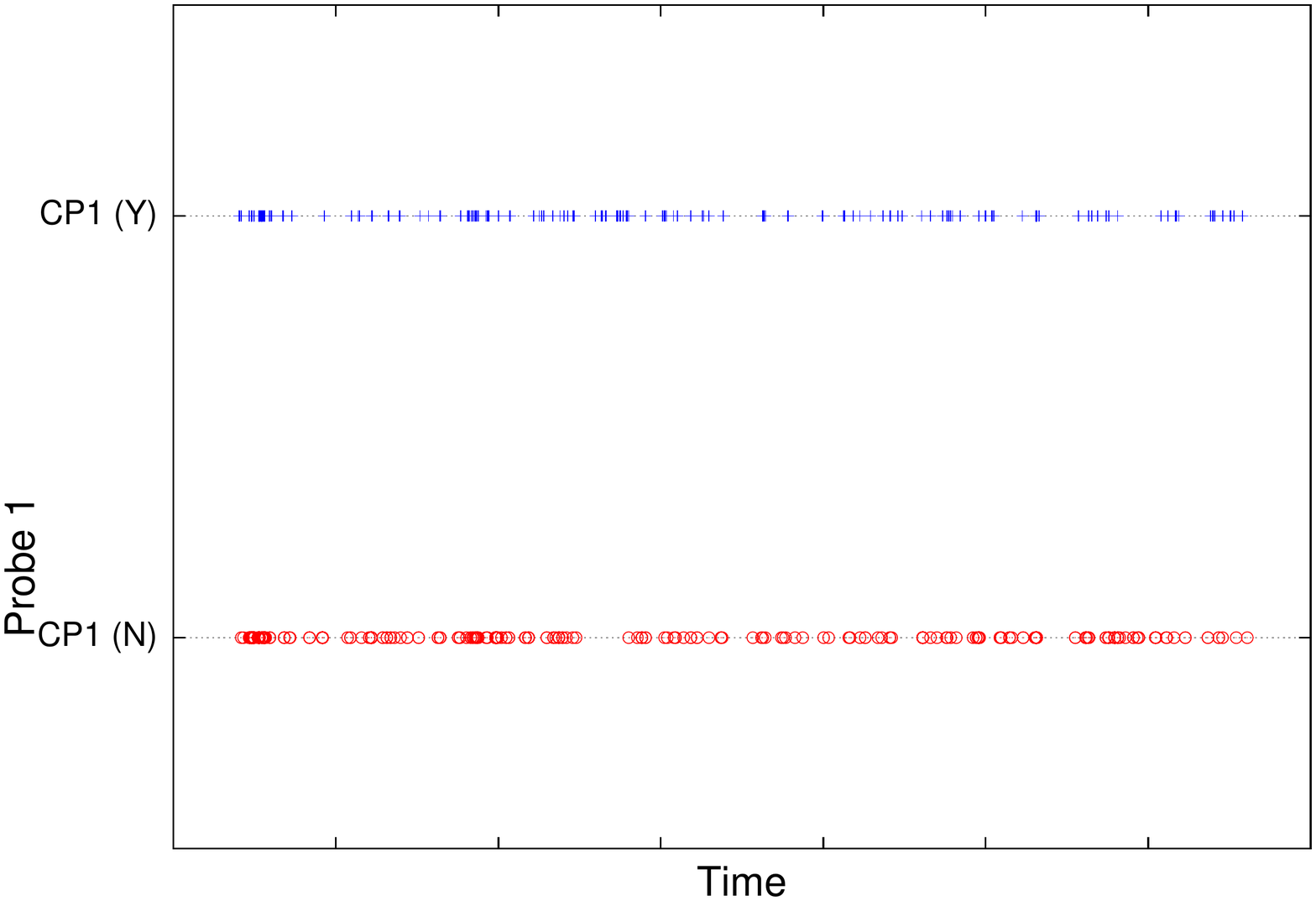}
   \includegraphics[width=\columnwidth]{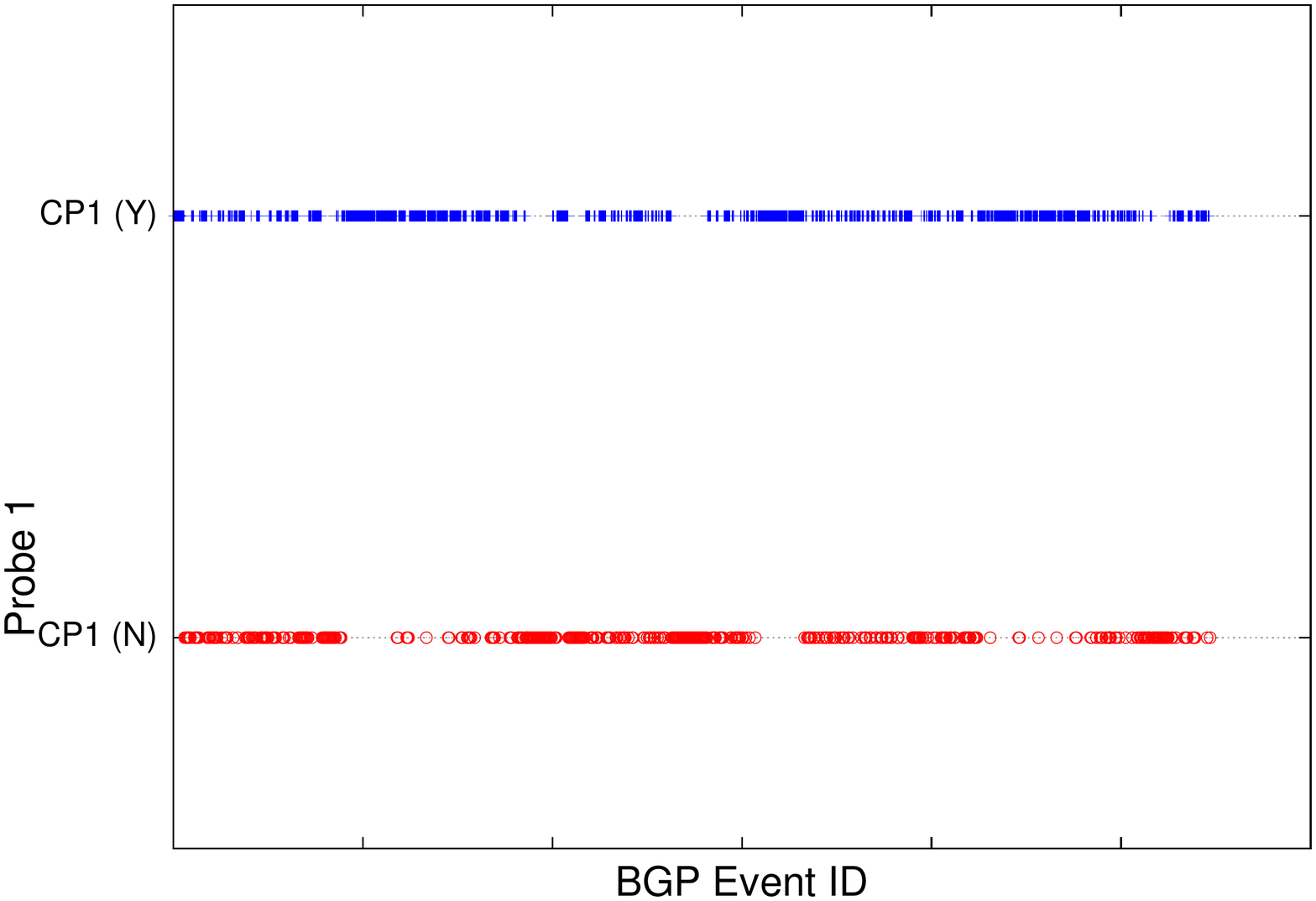}}
   \subfigure[AS 680, Measurement 1004 (target: 192.5.5.241)]{%
   \includegraphics[width=\columnwidth]{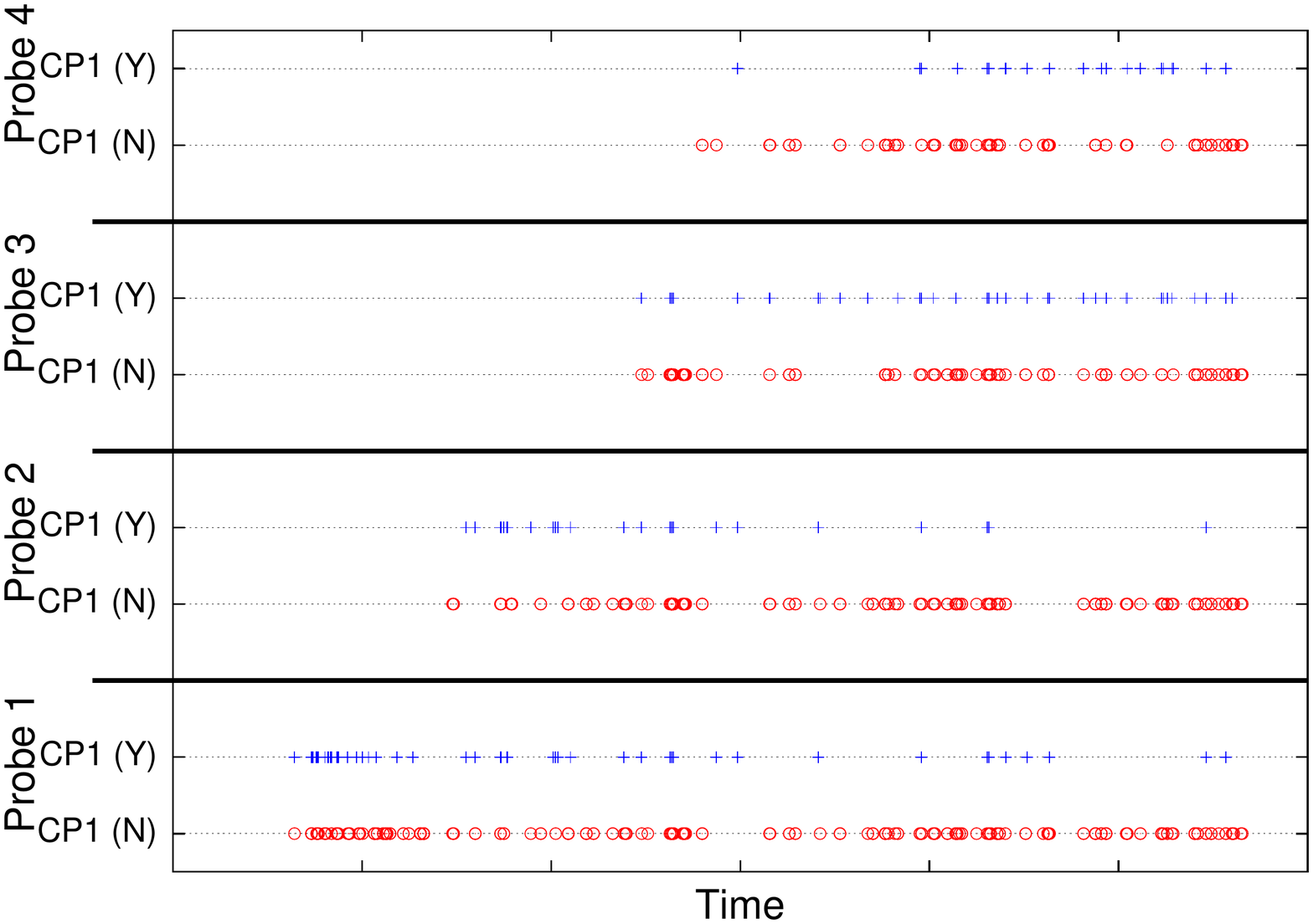}
   \includegraphics[width=\columnwidth]{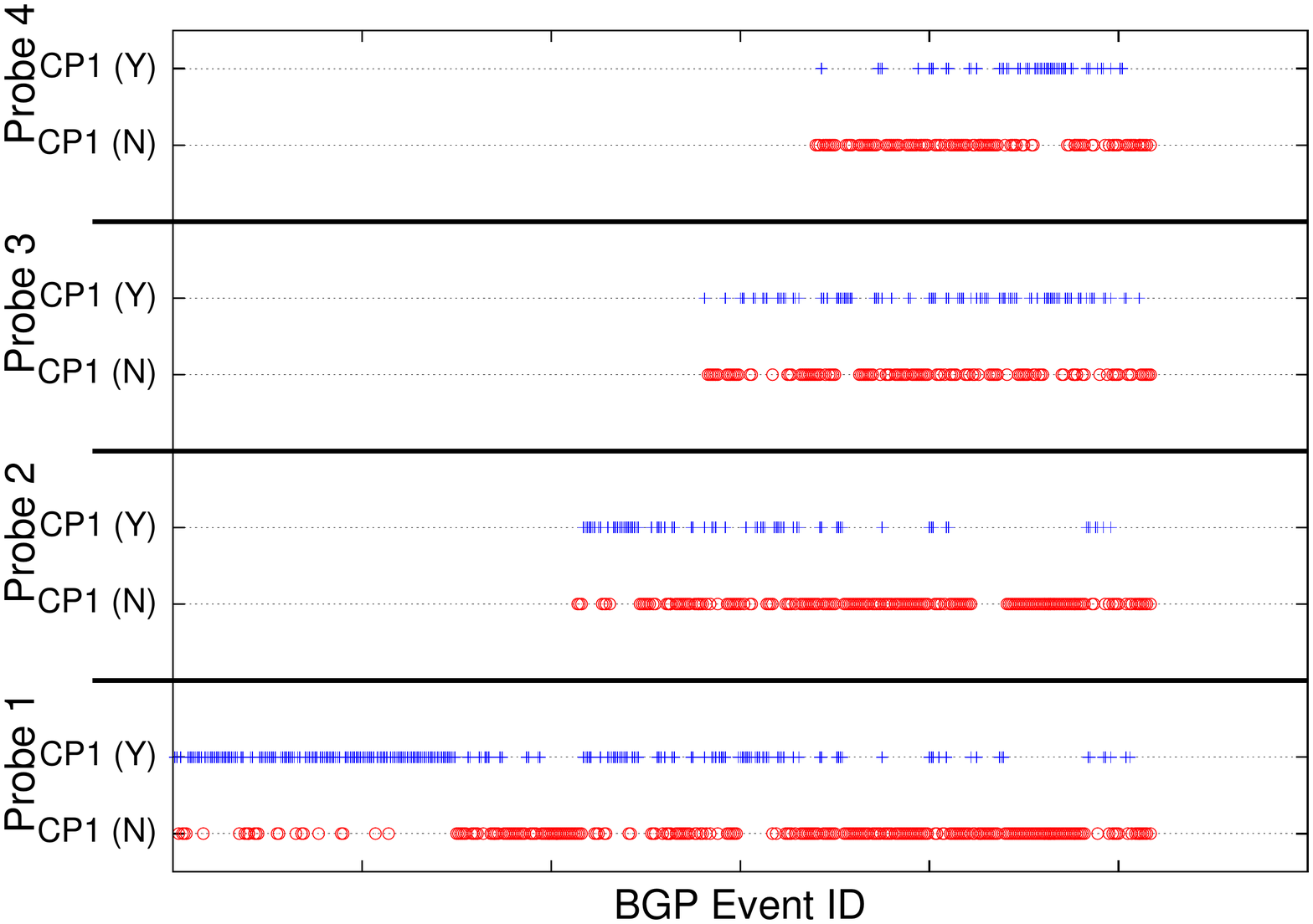}}
   \caption{Matching of the BGP updates seen by all the CPs of a specific AS with RTT values measured by all the probes of the same AS. Plots in each row are for a specific AS and measurement target. Each point in the plots is a BGP update. In left-side plots the X axis represents the time instant when BGP updates were observed, while in right-side plots it represents the sequence in which BGP updates occurred. In each plot, ordinates are divided into stripes, each corresponding to data from a specific probe. Each stripe contains two lines for each CP, representing matching (Y) and non-matching (N) BGP updates.}
   \label{fig:equivalenceclasses}
\end{figure*}

Besides considering aggregated correlation values, we further deepened the analysis by considering how single BGP events seen at distinct CPs match with RTT measurements recorded at different probes. A sample of the results of this analysis is in Fig.~\ref{fig:equivalenceclasses}. Plots in each row consider data from all the probes and CPs available within a specific AS, restricted to a specific measurement target but regardless of the prefix. Each point in the plots is a BGP update. The X axis represents timestamps of BGP updates (in left-side plots) or a progressive identifier assigned to these updates according to the order in which they occurred (in right-side plots). The Y axis is divided into stripes, one for each probe; within each stripe, two lines represent BGP updates seen at a specific CP and that matched (Y) or did not match (N) with an RTT changepoint recorded by the corresponding probe.
The topmost plots (a) show a good matching between a specific probe and CP, while other probes exhibit less correlation. This is even more evident by looking at the sequence of BGP updates in the right-side plot. Note that 2 CPs are actually available for this AS, but data for one of them has been suppressed in \textsc{Preprocessing} steps. The middle plots (b) show a good correlation between the probe and CP available within AS 513. Also in this plot only one of the 2 available CPs are visible. The bottommost plots (c) show how BGP updates recorded by a single CP may be more correlated with different probes depending on the time instant. Once again, this is more evident by looking at the sequence of updates. Also in this case, data from one additional CP was suppressed.

By looking at these plots, particularly (a) and (c), it is pretty evident that BGP updates from a specific CP may be more or less correlated with RTT values from a specific probe depending on the time instant. It is therefore possible to point out equivalence classes of CPs or probes, according to the behavior they exhibit over time. For example, in plot (c) probes 1 and 2 may be placed in the same equivalence class because they roughly matched the same BGP updates from the CP, whereas probes 3 and 4 may be placed in a different equivalence class because they matched other BGP updates.
This kind of analysis may help an operator in determining how the performance of specific network paths is affected by routing changes at specific BGP routers.

%% file: 055-validation.tex
\subsection{Validation with Traceroute Data}
\label{sec:experiments-validation}

The probes we take into account also perform traceroutes towards the same targets used for RTT measurements. This gives us the possibility to build a validation process for the correlation methodology described in Section~\ref{sec:methodology}. We explain here the validation process and the results we obtained using the data described in Section~\ref{sec:experiments-datasets}. Refer to Fig.~\ref{fig:methodology}(b) for an outline of the main steps described below.

The validation process applies to a single probe/CP pair, and considers a specific \textsf{Target} and a specific \textsf{Prefix}.
To perform the validation, we consider as first input a sequence of time-labeled traceroute measurements (performed with the standard \texttt{traceroute} command). Each measurement consists of a sequence of IP addresses and possibly includes ``null hops'', i.e., hosts that do not reply to packets sent by the probe. The second input is a partial result of the methodology in Section~\ref{sec:methodology}: it consists of the complete set of BGP updates collected by the selected CP for the \textsf{Prefix} of interest, where each BGP update is labeled as \emph{valid} if it has been retained after the \textsc{Preprocessing} step described in Section~\ref{sec:bgp-meth}, \emph{invalid} otherwise. The last input consists of the correlation factor for the considered probe/CP pair.

The collected data is subject to an \textsc{IP$\rightarrow$AS Mapping} step. The goal is to determine what ASes are traversed by each traceroute measurement, in order to match them with BGP routing changes happening at the same time. Methods for IP-to-AS mapping from the literature~\cite{mrwk-taaltt-03, mjrwk-saialfp-04, zowszbzz-fqputaltm-11} take into account many potential issues, e.g., the presence of IP addresses announced by Internet Exchange Points (IXPs) or peer ASes. However, all the existing approaches to the mapping rely on some preprocessing step and none of them is currently available as a public service or tool. In this paper we take a simple approach and retain the assumption supported by the literature that the IP-to-AS mapping derived from BGP routing tables is mostly correct. The main steps of our mapping step can be described as follows:
\begin{inparaenum}[1)]
   \item each private IP address at the beginning of the traceroute path is mapped to the AS of the probe originating the measurement;
   \item each remaining IP address is mapped to the most specific IP prefix containing it that is publicly announced on BGP and seen by RIS collectors. The first information is made available on the Atlas homepage~\cite{atlas}, while the latter can be retrieved using RIPEstat~\cite{ripestat}. For each prefix, the AS that announces it is elected as representative for all the IP addresses contained in the prefix. In case there is more than one AS announcing the same prefix, the one seen as the prefix originator by the majority of RIS collectors is elected. In case there is no IP prefix matching the IP address we map the latter to a special number representing an unknown AS;
   \item in the resulting sequence of ASes, identical consecutive AS numbers are collapsed;
   \item AS numbers corresponding to publicly known IXPs are removed from the sequence.
\end{inparaenum}

After performing a \textsc{Time alignment} step in the same way as for RTT measurements (see Section~\ref{sec:rtt-meth}),
%
we put together the traceroute measurements and the BGP routing changes in the \textsc{Matching} step. We consider the sequence of all BGP routing changes $u_{1}, \dots, u_{n}$ with related timestamps $t_{1}, \dots, t_{n}$. For each valid BGP routing change $u_{i}$ ($1<i<n$) we consider two time windows $T_{<} = [t_{i-1}, t_{i}]$ and $T_{>} = [t_{i}, t_{i+1}]$ which are determined as the stability periods before and after $u_{i}$. We then obtain $M_{<}$ and $M_{>}$, i.e. the two sequences of traceroute measurements respectively falling within $T_{<}$ and $T_{>}$. We discard valid BGP updates where either $M_{<}$ or $M_{>}$ are empty. Given the last traceroute measurement in $M_{<}$ with the highest timestamp we call $m_{i-1}$ the sequence of ASes obtained by mapping its IP addresses. Similarly, $m_{i}$ is the sequence of ASes corresponding to the first traceroute measurement in $M_{>}$. We combine the above information in a quadruple $q_{i} = (m_{i-1}, u_{i-1}, m_{i}, u_{i})$ and call $Q$ the set containing all such quadruples.

Finally, the output of the correlation between RTT measurements and BGP routing changes is validated as follows. The analysis is applied to each $q_{i} \in Q$ computed in the previous step and makes use of the outcome of the BGP-RTT \textsc{Matching} step for BGP routing change $u_{i}$ (see Section~\ref{sec:rtt-meth}). For each $q_{i} \in Q$ we simply check whether $u_{i-1} \neq u_{i}$ and $m_{i-1} \neq m_{i}$, and mark $u_i$ as ``validated'' if both conditions hold. Given such information, we can compute two quality measures for the \textsf{Validation} as follows. We first split $Q$ into two sets $Q_{+}$ and $Q_{-}$: the first contains all the $q_{i} \in Q$ such that the BGP routing change $u_{i}$ is correlated with an RTT measurement, while the second is defined as $Q_{-} = Q \setminus Q_{+}$. We then define the \emph{BGP-traceroute correlation factor} as the ratio $\frac{|{q_i \in Q_{+}:\ u_i \text{ is validated}}|}{|Q_{+}|}$ and the \emph{BGP-traceroute false negative factor} as the ratio $\frac{|{q_i \in Q_{-}:\ u_i \text{ is validated}}|}{|Q_{-}|}$. Intuitively, the first factor measures the precision of our BGP-RTT correlation methodology, while the second gives an indication of how many correspondences between the data plane and the control plane are not captured.

\begin{figure*}
   \centering
   \begin{tabular}{*{2}{m{5.8cm}}m{3cm}l}
      \includegraphics[width=.80\columnwidth]{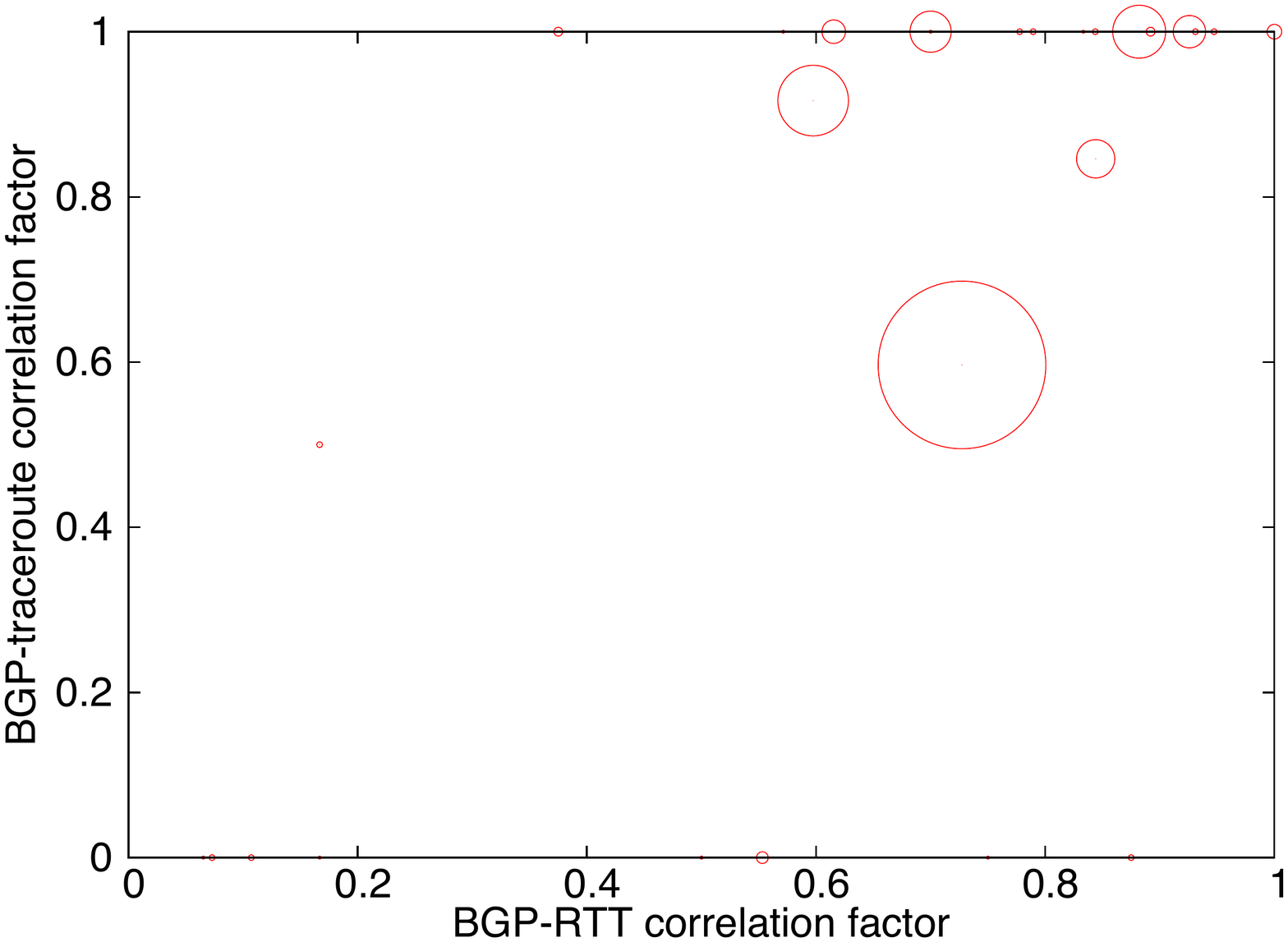} &
      \includegraphics[width=.80\columnwidth]{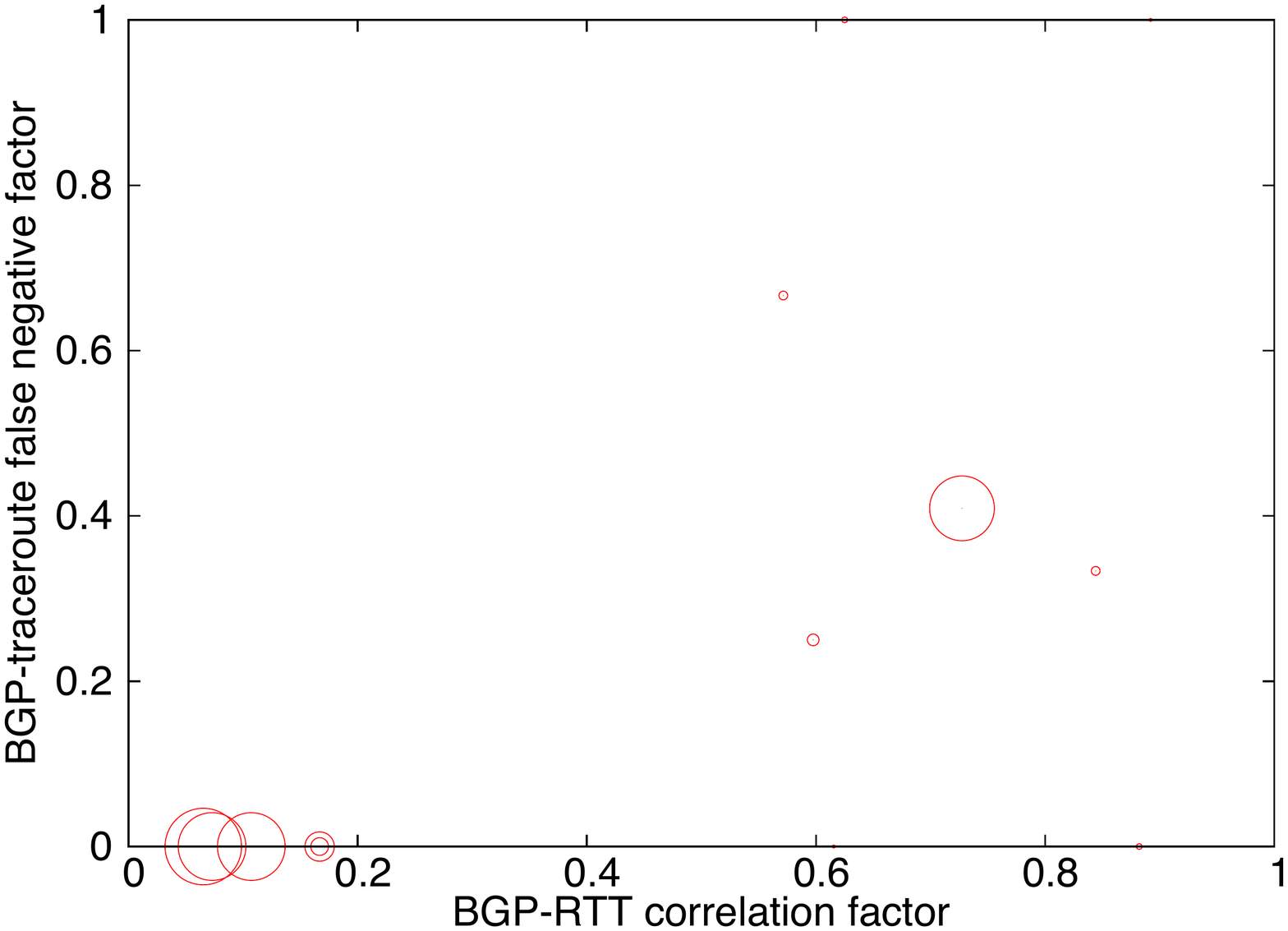} &
      \parbox{3.5cm}{Meas. 1001\\ (target: 193.0.14.129)\\(prefix: 193.0.14.0/24)}\\

      \includegraphics[width=.80\columnwidth]{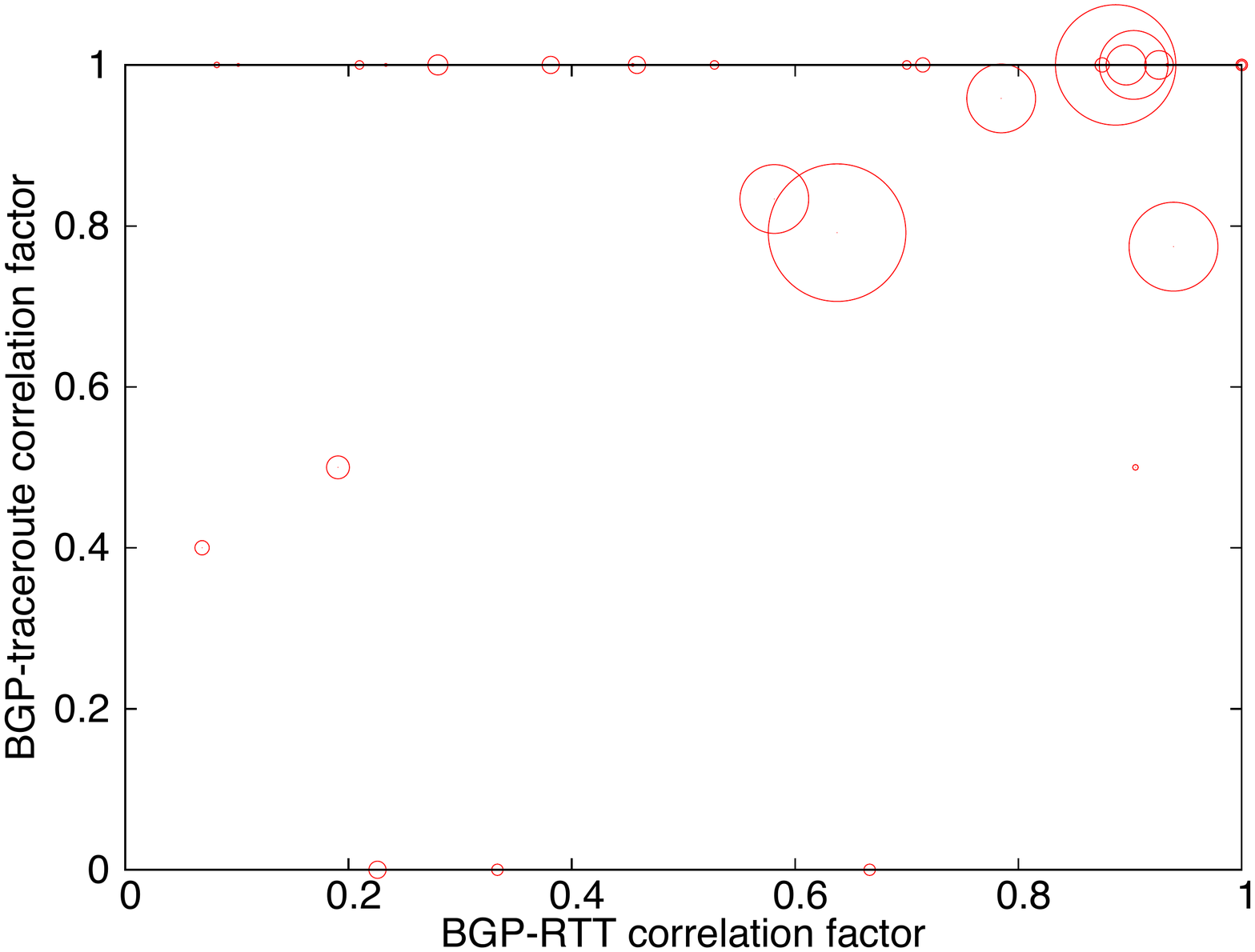} &
      \includegraphics[width=.80\columnwidth]{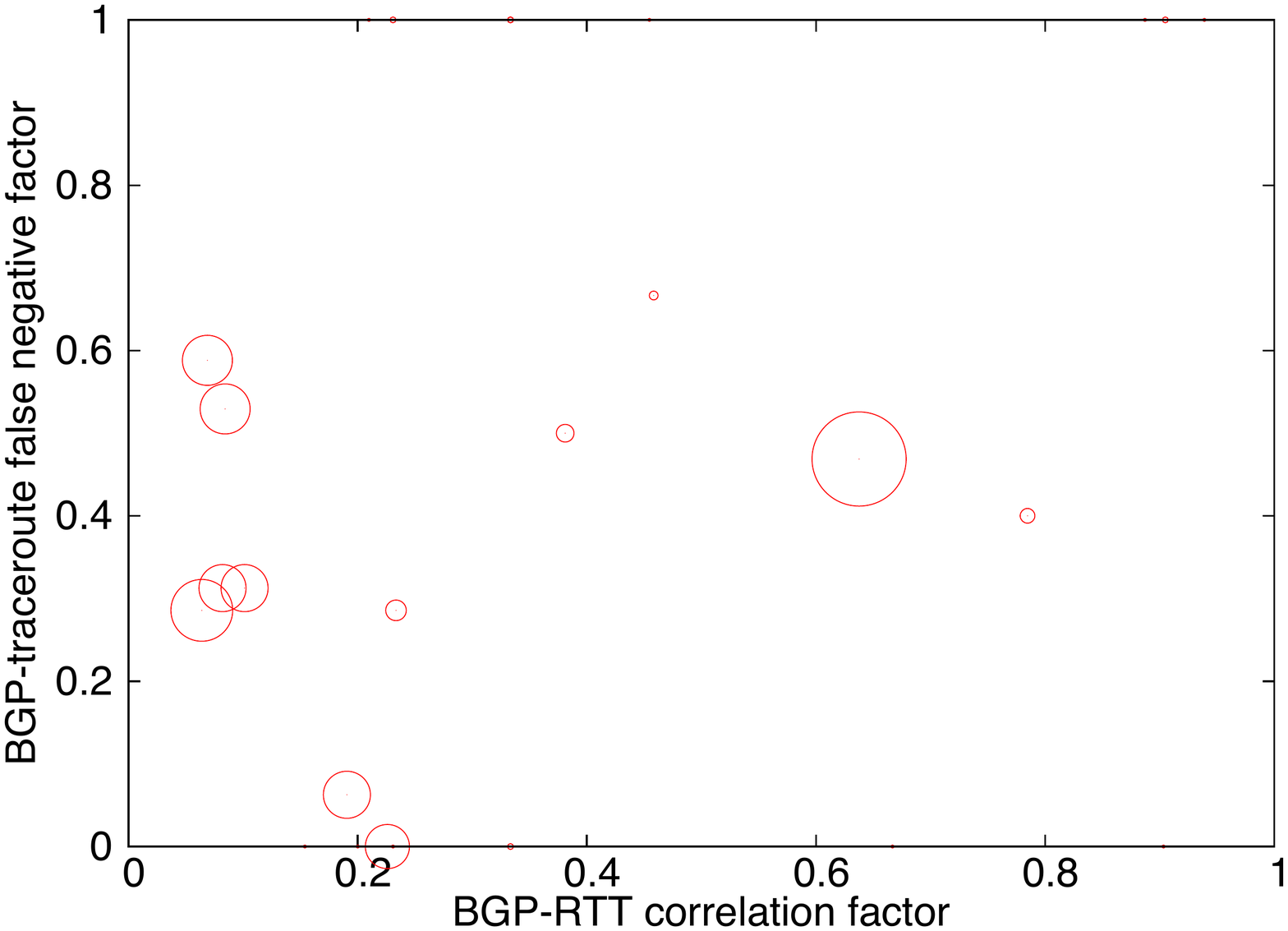} &
      \parbox{3.5cm}{Meas. 1003\\ (target: 193.0.0.193)\\(193.0.0.0/21)}\\
      
      \includegraphics[width=.80\columnwidth]{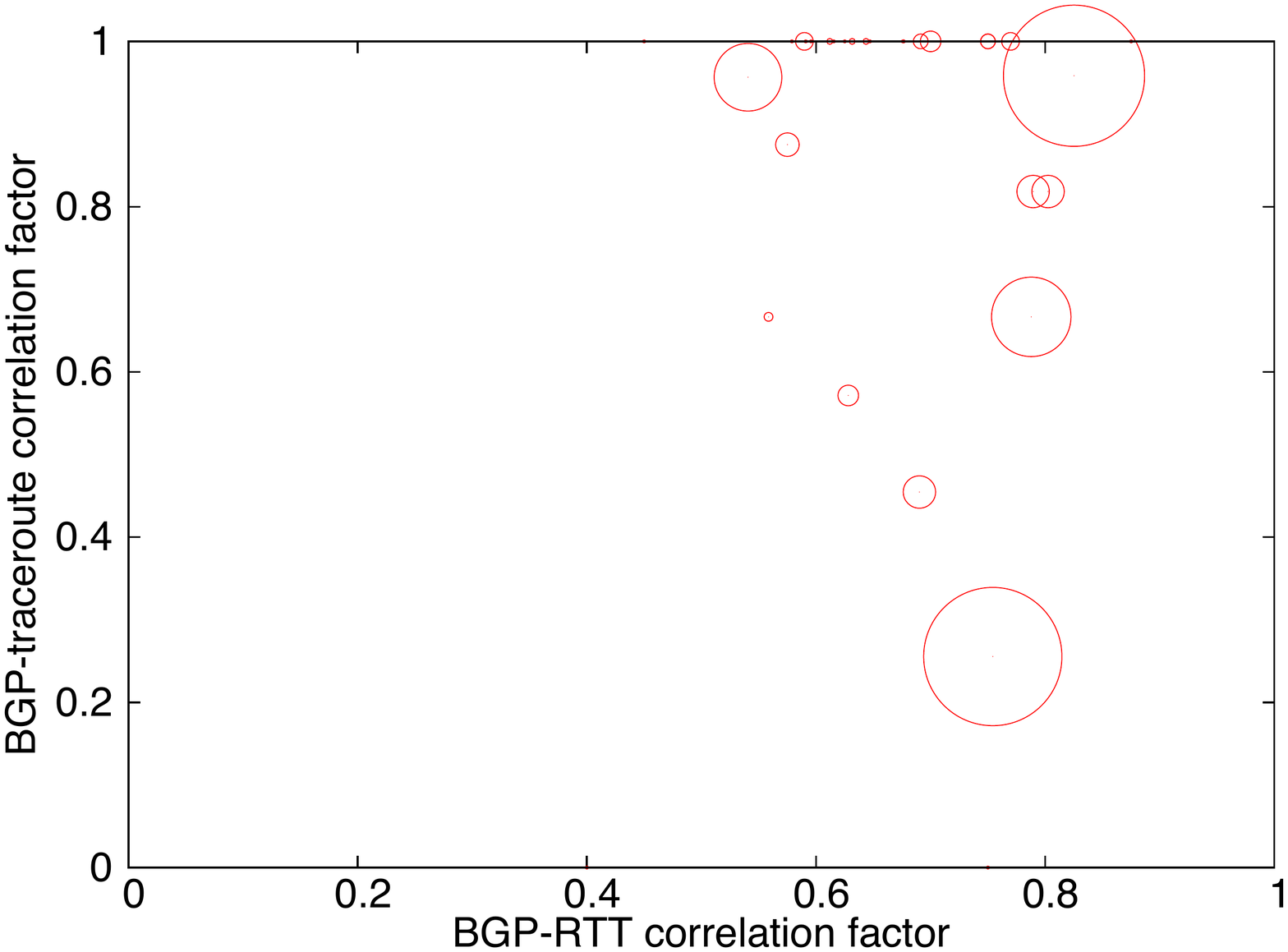} &
      \includegraphics[width=.80\columnwidth]{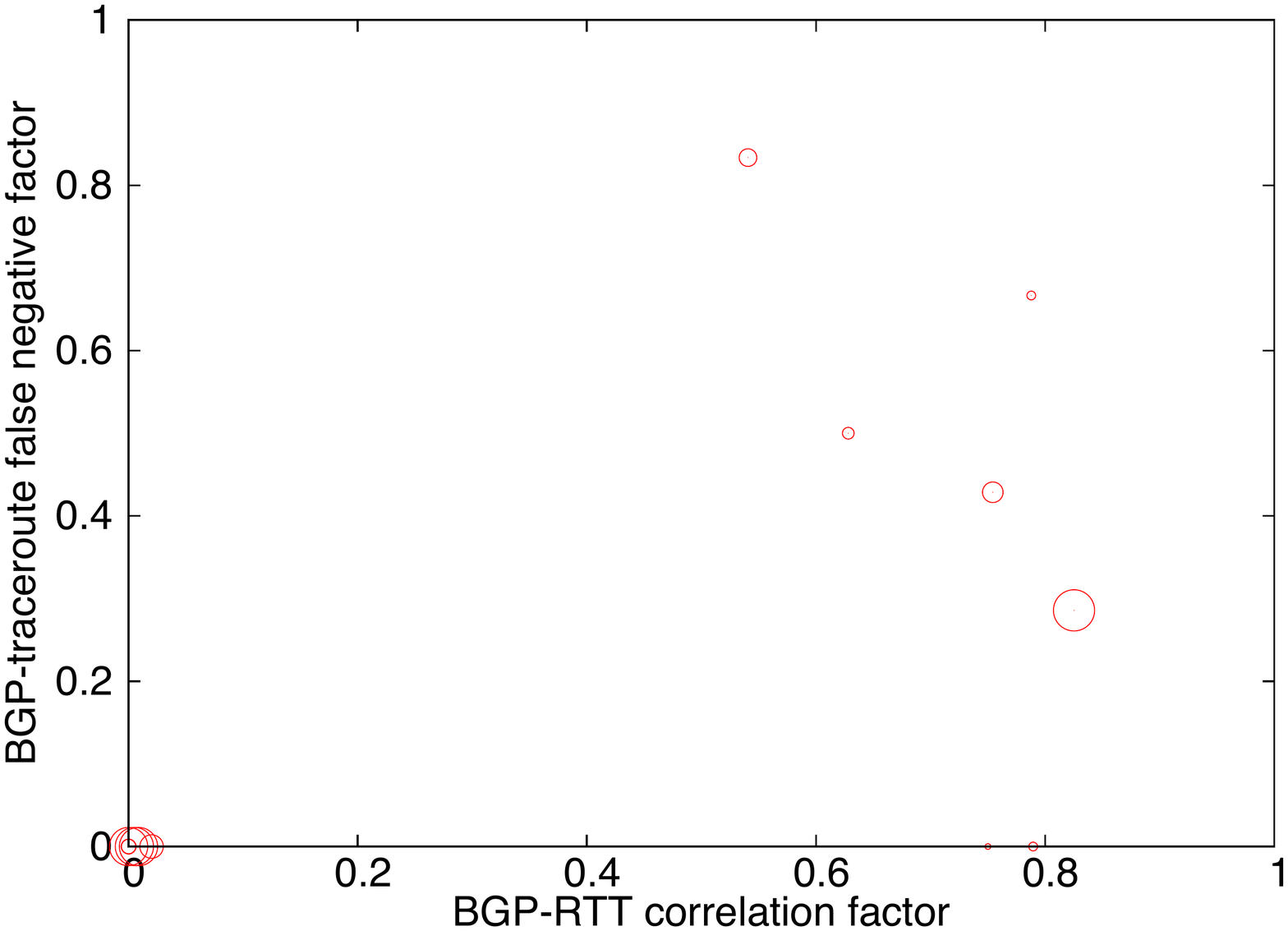} &
      \parbox{3.5cm}{Meas. 1004\\ (target: 192.5.5.241)\\(192.5.5.0/24)}\\
      
      \includegraphics[width=.80\columnwidth]{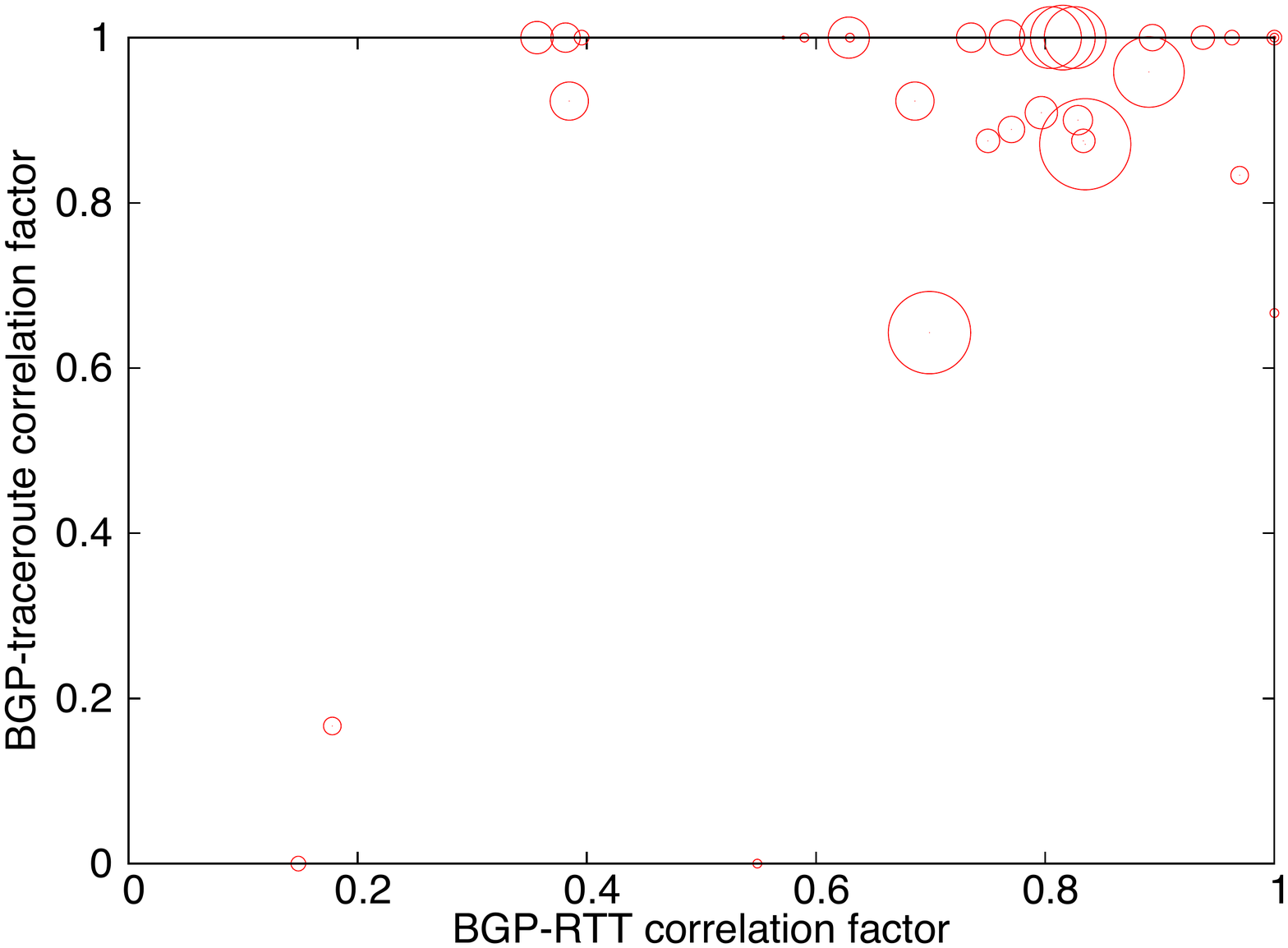} &
      \includegraphics[width=.80\columnwidth]{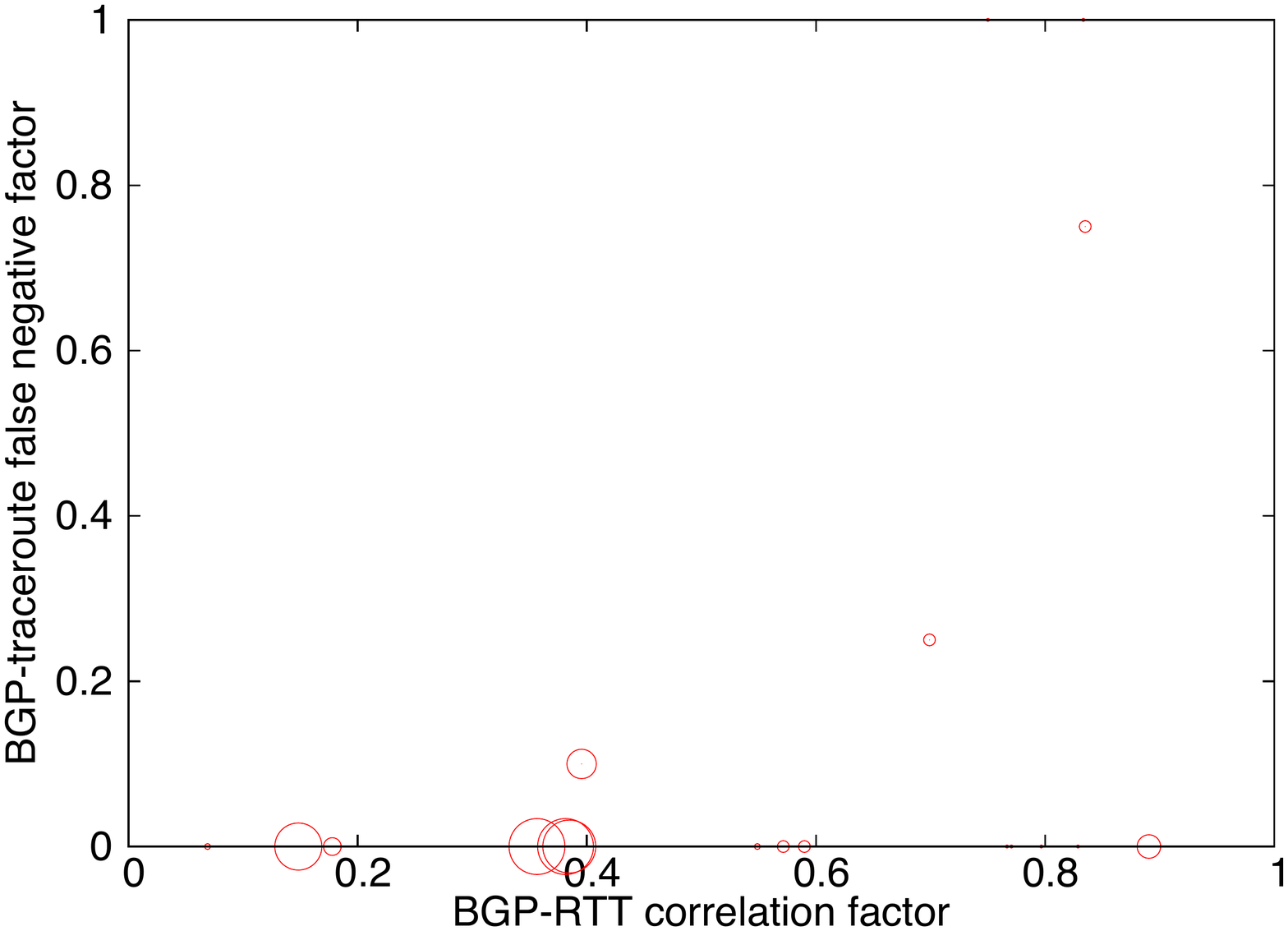} &
      \parbox{3.5cm}{Meas. 1005\\ (target: 192.36.148.17)\\(192.36.148.0/24)}\\
   \end{tabular}

   \caption{Validation with traceroute data. Circles represent probe/CP pairs. Each row contains two plots computed for each measurement. Left-side plots show the BGP-traceroute correlation factors and the BGP-RTT correlation factors for all the probe/CP pairs with $|Q_+|>0$. Right-side plots show the BGP-traceroute false negative factors and the BGP-RTT correlation factors for all the probe/CP pairs with $|Q_-|>0$. The size of each circle is proportional to $|Q_+|$ and $|Q_-|$, respectively.}
   \label{fig:bgp-traceroute-validation}
\end{figure*}
%

The results of the validation are in Fig.~\ref{fig:bgp-traceroute-validation}. Each row contains plots related to a specific measurement, thus determining both the \textsf{Target} and the \textsf{Prefix}. Circles in each plot represent probe/CP pairs. Other features are described in the figure caption.

The left-side plots exhibit common features that confirm the precision of our methodology. In particular, the vast majority of probe/CP pairs with a high BGP-RTT correlation factor (greater than $0.6$) have a high BGP-traceroute correlation factor. That means that BGP routing changes that are well correlated with RTT changepoints are also well correlated with traceroutes. The data for measurement $1004$ apparently contains some false positives: by manual inspection we found out that they are simply due to some failed IP-to-AS mappings.

The right-side plots help us understand how many correlations cannot be detected with our methodology. In these plots we expect to see BGP-RTT false negative factors close to zero for probe/CP pairs whose BGP-RTT correlation factor is low.
Namely, poorly correlated BGP and RTT data should find no evidence of correlation even at the traceroute level.
Measurements $1001$, $1004$, and $1005$ match our expectation, while $1003$ is the only notable exception. A manual inspection of traceroute data confirmed the correlation with BGP routing changes. After comparing several metrics on the input data (e.g. mean value and variance of RTT measurements, number of computed changepoints, etc.), we suspect that this exception may be related to the fact that both the unicast target and most of the probes are topologically close, leading to shorter AS-paths and, possibly, less identifiable RTT changes.

%% file: main.bbl
\begin{thebibliography}{10}

\bibitem{samknows}
\url{http://www.samknows.com/broadband/}.

\bibitem{agcom}
\url{https://www.misurainternet.it/}.

\bibitem{ark}
\url{http://www.caida.org/projects/ark/}.

\bibitem{mlab}
\url{http://www.measurementlab.net/}.

\bibitem{r-changepoint-package}
\url{http://cran.r-project.org/web/packages/changepoint/}.

\bibitem{atlas}
{RIPE Atlas}.
\newblock \url{http://atlas.ripe.net/}.

\bibitem{ripestat}
{RIPEstat}.
\newblock \url{https://stat.ripe.net/}.

\bibitem{bn-dacta-93}
M.~Basseville and I.V. Nikiforov.
\newblock Detection of abrupt changes: Theory and application, '93.

\bibitem{dds-vdcbbrrtdam-13}
G.~{Da Lozzo}, G.~{Di Battista}, and C.~Squarcella.
\newblock Visual discovery of the correlation between {BGP} routing and
  round-trip delay active measurements.
\newblock {\em Computing}, pages 1--11, 2013.

\bibitem{hm-owdmc-07}
A.~Hernandez and E.~Magana.
\newblock One-way delay measurement and characterization.
\newblock In {\em Proc. ICNS}, 2007.

\bibitem{1381461}
B.~Jackson, J.D. Scargle, D.~Barnes, S.~Arabhi, A.~Alt, P.~Gioumousis, E.~Gwin,
  P.~Sangtrakulcharoen, L.~Tan, and Tun~Tao Tsai.
\newblock An algorithm for optimal partitioning of data on an interval.
\newblock {\em Signal Processing Letters}, 12(2):105--108, 2005.

\bibitem{SMJ:SMJ819}
D.J. Ketchen and C.L. Shook.
\newblock The application of cluster analysis in strategic management research:
  an analysis and critique.
\newblock {\em Strategic Management Journal}, 17(6):441--458, 1996.

\bibitem{kfe-odcwlcc-12}
R.~Killick, P.~Fearnhead, and I.A. Eckley.
\newblock Optimal detection of changepoints with a linear computational cost.
\newblock {\em Journal of the American Statistical Association},
  107(500):1590--1598, 2012.

\bibitem{Lee:2012:MSA:2398776.2398788}
M.~Lee, N.~Duffield, and R.R. Kompella.
\newblock {MAPLE}: A scalable architecture for maintaining packet latency
  measurements.
\newblock In {\em Proc. IMC}, 2012.

\bibitem{mgwyzehs-rdmncsp-11}
A.~Mahimkar, Z.~Ge, J.~Wang, J.~Yates, Y.~Zhang, J.~Emmons, B.~Huntley, and
  M.~Stockert.
\newblock Rapid detection of maintenance induced changes in service
  performance.
\newblock In {\em Proc. CoNEXT}, 2011.

\bibitem{msgswyze-dpiulon-10}
A.A. Mahimkar, H.H. Song, Z.~Ge, A.~Shaikh, J.~Wang, J.~Yates, Y.~Zhang, and
  J.~Emmons.
\newblock Detecting the performance impact of upgrades in large operational
  networks.
\newblock In {\em Proc. SIGCOMM}, 2010.

\bibitem{mjrwk-saialfp-04}
Z.M. Mao, D.~Johnson, J.~Rexford, J.~Wang, and R.~Katz.
\newblock Scalable and accurate identification of {AS}-level forwarding paths.
\newblock In {\em Proc. INFOCOM}, 2004.

\bibitem{mrwk-taaltt-03}
Z.M. Mao, J.~Rexford, J.~Wang, and R.H. Katz.
\newblock Towards an accurate {AS}-level traceroute tool.
\newblock In {\em Proc. SIGCOMM}, 2003.

\bibitem{5542733}
E.S. Myakotnykh, B.E. Helvik, O.J. Wittner, O.~Kvittem, J.K. Hellan,
  T.~Skjesol, and A.~Oslebo.
\newblock Measurement and analysis of end-to-end dependability characteristics
  in the global network.
\newblock In {\em Proc. IWQoS}, 2010.

\bibitem{pzmh-undcre-07}
H.~Pucha, Y.~Zhang, Z.M. Mao, and Y.C. Hu.
\newblock Understanding network delay changes caused by routing events.
\newblock In {\em SIGMETRICS}, 2007.

\bibitem{ris}
{RIPE NCC}.
\newblock {Routing Information Service (RIS)}.
\newblock
  \url{http://www.ripe.net/data-tools/stats/ris/routing-information-service}.

\bibitem{bismark}
S.~Sundaresan, W.~de~Donato, N.~Feamster, R.~Teixeira, S.~Crawford, and
  A.~Pescap\`e.
\newblock Broadband internet performance: A view from the gateway.
\newblock In {\em Proc. SIGCOMM}, 2011.

\bibitem{route}
{University of Oregon}.
\newblock {Route Views Project}.
\newblock \url{http://www.routeviews.org/}.

\bibitem{zmz-edrdes-08}
Y.~Zhang, Z.M. Mao, and M.~Zhang.
\newblock Effective diagnosis of routing disruptions from end systems.
\newblock In {\em Proc. NSDI}, 2008.

\bibitem{zowszbzz-fqputaltm-11}
Y.~Zhang, R.~Oliveira, Y.~Wang, S.~Su, B.~Zhang, J.~Bi, H.~Zhang, and L.~Zhang.
\newblock A framework to quantify the pitfalls of using traceroute in
  {AS}-level topology measurement.
\newblock {\em Jour. Sel. Areas Comm.}, 29(9):1822--1836, 2011.

\end{thebibliography}
